\documentclass[aos,preprint]{imsart}

\RequirePackage[OT1]{fontenc}
\RequirePackage{amsthm,amsmath}
\RequirePackage{natbib}
\RequirePackage[colorlinks,citecolor=blue,urlcolor=blue]{hyperref}

\usepackage{xr}
\makeatletter
\newcommand*{\addFileDependency}[1]{
	\typeout{(#1)}
	\@addtofilelist{#1}
	\IfFileExists{#1}{}{\typeout{No file #1.}}
}
\makeatother

\newcommand*{\myexternaldocument}[2]{%
	\externaldocument[#2]{#1}%
	\addFileDependency{#1.tex}%
	\addFileDependency{#1.aux}%
}
\myexternaldocument{sup-paper07122017}{}
\usepackage{amsthm}
\usepackage{pdfsync}
\usepackage{color}
\usepackage{multirow}
\usepackage{graphicx}
\usepackage{amsmath, amssymb, amsfonts, amsthm}
\usepackage{epstopdf}
\usepackage{graphics}
\usepackage{color}

\usepackage{caption}
\usepackage{subcaption}
\usepackage{mathtools}
\newcommand{\field}[1]{\mathbb{#1}}
\newcommand{\R}{\field{R}}

\newcommand{\p}{\field{P}}
\newcommand{\N}{\field{N}}

\newcommand{\E}{\field{E}}

\newcommand{\FF}{\mathcal{F}}

\newcommand{\convd}{\overset{\mathcal{D}}{\Longrightarrow}}

\def\qed{\hfill$\diamondsuit$}

\def\argmin{\mathop{\mbox{argmin}}}
\theoremstyle{Conjecture}
\theoremstyle{example}
\theoremstyle{remark}
\theoremstyle{lemma}
\theoremstyle{definition}
\theoremstyle{corol}
\theoremstyle{proposition}
\theoremstyle{condition}
\newtheorem{theorem}{Theorem}[section]

\newtheorem{remark}{Remark}[section]
\newtheorem{lemma}{Lemma}[section]
\newtheorem{definition}{Definition}[section]

\newtheorem{proposition}{Proposition}[section]

\newtheorem{assumption}{Assumption}[section]

\newtheorem{algorithm}{Algorithm}[section]

\def\lf{\lfloor}
\def\rf{\rfloor}

\newcommand{\bea}{\begin{eqnarray*}}
	\newcommand{\eea}{\end{eqnarray*}}
\newcommand{\be}{\begin{eqnarray}}
\newcommand{\ee}{\end{eqnarray}}

\usepackage{booktabs}
\usepackage{hyperref}

\startlocaldefs
\numberwithin{equation}{section}
\theoremstyle{plain}

\endlocaldefs

\begin{document}
	
	\begin{frontmatter}
        		\title{Detecting relevant changes in the mean of  non-stationary processes - a mass excess approach}
		\runtitle{Relevant changes via a mass excess}

	\begin{aug}               \author{\fnms{Holger} \snm{Dette}\thanksref{m1}\ead[label=e1]{holger.dette@ruhr-uni-bochum.de}},
		\author{\fnms{Weichi} \snm{Wu}\thanksref{m1}\thanksref{m2}\ead[label=e2]{wuweichi@mail.tsinghua.edu.cn}}
		
		
		\affiliation{Ruhr-Universit\"at Bochum \thanksmark{m1}
			and Tsinghua University\thanksmark{m2}}
		\address{Ruhr University Bochum\\
			Building IB 2/65
			Universit\"{a}tsstrasse 150,\\
			D-44801 Bochum,\\
			Germany\\
			\printead{e1}\\
			\phantom{E-mail:\ }}
		\address{Center for Statistical Sciences,\\
			Department of Industrial Engineering,\\
			Tsinghua University, Beijing, \\China\\\printead{e2}
			\phantom{E-mail:\ }}

	\end{aug}

		\begin{abstract}
			This paper considers the problem of testing if a sequence of means $(\mu_t)_{t =1,\ldots ,n }$ of a non-stationary time series  $(X_t)_{t =1,\ldots ,n }$  is stable in the sense that the difference of the means $\mu_1$  and $\mu_t$ between the initial time $t=1$ and any  other time is smaller than a given threshold, that is $ | \mu_1  - \mu_t | \leq c $ for  all $t =1,\ldots ,n $. A test for hypotheses of this type is developed using a bias corrected
			monotone rearranged local linear estimator and asymptotic normality of the corresponding test statistic is established.  As the asymptotic variance depends on the location of the roots of the equation $| \mu_1  - \mu_t | = c$ a new bootstrap procedure is
			proposed to obtain critical values and its consistency is established. As a consequence we are able to  quantitatively describe relevant deviations of a non-stationary sequence from its initial value.  
			The results are illustrated by means of a simulation study and by analyzing data examples. 
		\end{abstract}

		
		
	\end{frontmatter}

\section{Introduction} \label{sec1}
\def\theequation{1.\arabic{equation}}
\setcounter{equation}{0}

A frequent problem in  time series analysis is the detection
 of structural breaks. Since the pioneering work of  \cite{page1954} in quality control
  change point detection  has become an important tool with numerous applications in  economics, climatology, engineering, hydrology 
and many authors have developed statistical tests  for  the problem of detecting structural breaks
or change-points in various models. Exemplarily we mention    \citet{chow:1960}, \citet{brown:1975}, \citet{ploberger:1988}, \cite{andrews1993}, \cite{baiper1998} and  \cite{aueetal2009}] and refer to the  work of \cite{auehor2013} and \cite{Jandhyala2013} for more recent reviews.

Most of the literature on testing for structural breaks formulates the hypotheses
such that in the statistical model the stochastic process under the null hypothesis of  ``no change-point'' is stationary. For example, in the problem of testing if a
 sequence of means $(\mu_t)_{t =1,\ldots ,n }$ of a non-stationary time series  $(X_t)_{t =1,\ldots ,n }$ is stable  it is often assumed that
 $X_t =  \mu_t  + \varepsilon_t$ with a stationary error process $( \varepsilon_t)_{t =1,\ldots ,n }$. The null hypothesis
 is then given by
 \begin{align} \label{null}
H_0: \mu_{1} = \mu_{2} = \dots = \mu_{n},
\end{align}
while the  alternative  (in the simplest case  of only one structural break) is defined  as
 \begin{align} \label{nulla}
H_1:
\mu_{(1)}  =
\mu_{1} = \mu_{2} = \dots = \mu_{k} \ \neq \ \mu_{k+1}  = \mu_{k+2} = \dots = \mu_n =\mu_{(2)},
\end{align}
where $k \in \{ 1,\dots,n \}$ denotes the (unknown) location of the change.
 The formulation of the null hypothesis in the form \eqref{null} facilitates the analysis of the distributional properties of a corresponding test statistic substantially, because one can work under the assumption
 of stationarity. Consequently, it is a  very useful  assumption   from a theoretical point of view.

 On the other hand, if
the differences $ \{ |\mu_1 - \mu_t| \}_{t=2,\ldots , n}$  are   rather ``small'',
a modification of the statistical analysis might not be necessary although  the test rejects the ``classical''
null hypothesis \eqref{null} and detects non-stationarity.
 For example,  as pointed out by \cite{DW2016},
in risk management one wants to  fit  a  model for forecasting the Value at Risk from ``uncontaminated data'', that means from data after the last change-point. If the  changes are small they might not yield large changes in the Value at Risk.  Now using only the uncontaminated data might
decrease the bias but increases the variance of a prediction. Thus, if the changes are small,  the forecasting quality might not necessarily decrease and -
in the best case - would only improve slightly. Moreover, any benefit with respect to statistical accuracy could be negatively overcompensated by additional transaction costs.

In order to address  these issues \cite{DW2016} proposed to investigate  {\it precise} hypotheses in the context of change point analysis, where one does not
test for exact equality, but only looks for    ``similarity'' or a ``relevant'' difference. This concept is well known in biostatistics [see, for example,  \cite{wellek2010testing}] but has also been used to investigate the similarity of distribution functions  [see \cite{cuevasetal2008,?lvarez-esteban2012}
among others]. 
In the  context of detecting a 
change in a  sequence of means (or other parameters of the marginal distribution)  \cite{DW2016}
assumed two stationary phases and tested if 
 the difference before and after the change point 
is small, that is 
\begin{align} \label{nullb}
H_0 : | \mu_{(1)}   - \mu_{(2)}  | \leq c 
~\mbox{ versus } ~H_1:| \mu_{(1)}   - \mu_{(2)}  | >  c,
\end{align}
where $c>0$ is a given constant specified by the concrete application (in the example of the previous paragraph $c$ could be determined
by the transaction costs).
Their approach heavily relies on the fact
that the process before and after the change point is stationary, but this assumption might also be
questionable in many  applications.

A similar idea can be used  to specify the economic design of control charts for quality control purposes. While in
change-point analysis the focus is on testing for the presence of a 
change and on estimating the time at which a change  occurs once it  has been detected, control charting has typically been focused more on detecting such a change as quickly as possible after it occurs {[see for example \cite{champwood1987}, \cite{woodmont1999} among many others]. In particular control charts are related to sequential change point detection, while the focus of the cited literature is on retrospective change point detection.}

\begin{figure}[t]
	\centering
	\includegraphics[width=7cm,height=5cm]{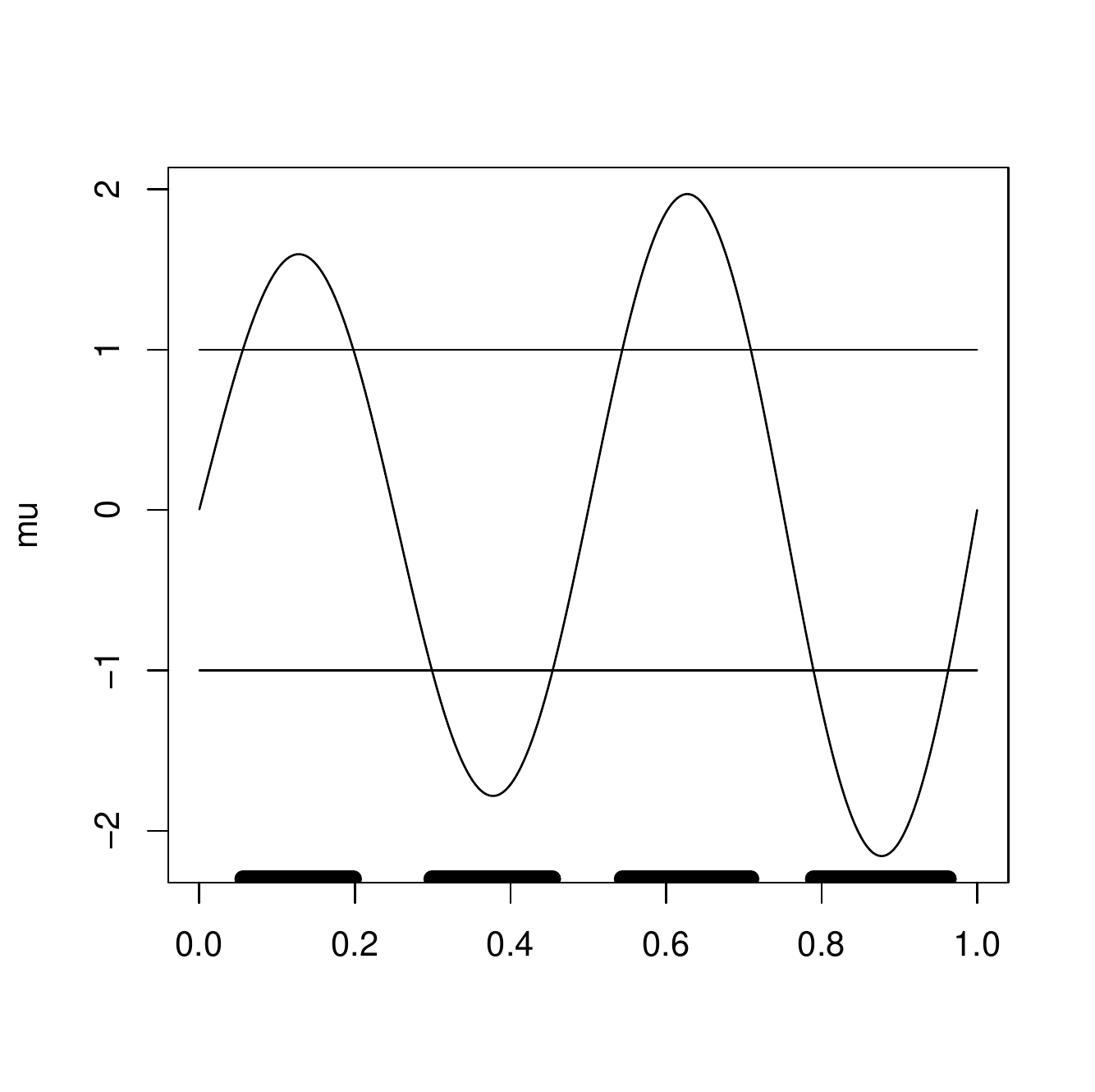}
	\vspace{-.4cm}
	\caption{\it Illustration of the set ${\cal M}_c$ in \eqref{mc}.}
	\label{illustration}
\end{figure}

{
 In the present paper  we investigate alternative relevant
 hypotheses in the retrospective change point problem,} which are  motivated by the  observation that   in many applications
 the assumption of two stationary phases 
(such as constant means before and after the change point) cannot be justified as   the process parameters
 change continuously in time. 
 For this purpose we consider the location scale model
 \begin{align}\label{protype}
X_{i,n} =\mu(i/n)+\epsilon_{i,n},
\end{align}
 where $\{ \epsilon_{i,n}  \colon  i=1,\ldots , n \}_{n\in \N} $ denotes a triangular  array of  centered
random variables (note that we do not assume that the ``rows'' $ \{ \epsilon_{j,n}: j=1,\ldots , n  \} $ are stationary)  and
$\mu : [0,1] \to \R$ is the unknown mean function.
 We define a change as {\it relevant}, if the amount of the change and the time period where the change occurs are reasonably large.
More precisely, for a level $c >0 $   we consider the {\it level set}
\begin{align} 
\label{mc}
  {\cal M}_c  =  \{ t \in [0,1] \colon   | \mu (t) -  \mu ({0}) | > c \}
\end{align}
of all points $t \in [0,1]$, where the mean function differs from its  original value at the point $0$ by an amount larger than $c$. 
The situation is illustrated in Figure \ref{illustration}, where
the curve represents the mean function  $\mu $ with $\mu(0)=0$ and
the lines in boldface represent the set ${\cal M}_c$ (with $c=1$).
These periods resemble in some sense popular run rules from the statistical quality control literature which signal if $k$ of  the  last  $m$ standardized  sample  means  fall  in  the interval [see for example
\cite{champwood1987}].

Define
 \begin{align} \label{tc}
 T_c :=  \lambda (  {\cal M} _c )
\end{align}
as  the corresponding {\it excess} measure, where $ \lambda  $ denotes the Lebesgue measure. 
We now propose to investigate the hypothesis that the  relative time, where  this difference is  larger than $c$  does not exceed
a given constant, say  $\Delta \in (0,1)$, that is
 \begin{align}
 \label{nullnew}
H_0 : T_c    \leq \Delta ~~\mbox{versus} ~~ H_1:   T_c     > \Delta ~.
\end{align}

We consider the change as {\it relevant}, if the Lebesgue measure $ T_c =  \lambda (  {\cal M} _c ) $ is larger than the threshold $ \Delta$. 
Note that this includes the case when a change (greater than $c$) occurs at some point $t_1 <1- \Delta$ and the mean level remains constant otherwise.

In many applications it might also be of interest to
investigate one-sided hypotheses, because one wants to detect a change in certain direction. For this purpose  we also consider the sets ${\cal M}_c^\pm
= \{ t \in [0,1] \colon   \pm ( \mu (t) -  \mu ({0}) ) > c \} $
and define the hypotheses
 \begin{align}
 \label{nullnew+}
H_0^+ : T_c^+ = \lambda (  {\cal M}_c^+ )    \leq \Delta ~~\mbox{versus} ~~ H_1^+:   T_c^+      > \Delta ~,  \\
H_0^- : T_c^- = \lambda (  {\cal M}_c^- )    \leq \Delta ~~\mbox{versus} ~~ H_1^-:   T_c^-      > \Delta ~.
 \label{nullnew-}
\end{align}
The hypotheses \eqref{nullnew}, \eqref{nullnew+} and \eqref{nullnew-} require the  specification of two parameters $\Delta$ and $c$ and in  a concrete application  both parameters have to be defined after a careful discussion with the practitioners.
In particular  they will be different in different fields of application. Another possibility is to investigate a relative deviation from the mean, that is: $\mu (t) $ deviates from $\mu (0)$ relative 
to $\mu (0)$ by at most $x\%$ (see Section \ref{sec222} for a discussion of this measure). 

Although  the mean function  in model \eqref{protype}
cannot be assumed to be monotone, we
use  a  monotone rearrangement type estimator
[see  \cite{DNP2006}] to estimate  the quantities $T_c$, $T_c^+$, $T_c^-$, and  propose to reject the null hypothesis \eqref{nullnew}, \eqref{nullnew+} \eqref{nullnew-}   for large values of the corresponding test statistic.
We  study  the properties of these estimators and  the  resulting tests
  in   a model of the form  \eqref{protype} with  a locally stationary error process,  which have found considerable interest in the literature [see
 \cite{DA1997}, \cite{NVK2000}, \cite{OVG2005}, \cite{zhou2009local} and   \cite{VO2012} 
   among others].  In particular we do {\bf not} assume that  the  underlying process is stationary, as the mean function
 can vary smoothly in time and the error process is non-stationary. Moreover, we also allow 
 that the derivative of the mean  function $\mu$ may vanish  on the set  of {\it critical roots}  
 $${\cal C} = \{t \in [0,1]
 \colon  |\mu (t) - \mu (0) | =  c \} $$
  and prove that  appropriately standardized versions of the monotone rearrangement estimators are
consistent for $T_c$, $T_c^+$ and $T_c^-$, and asymptotically normally distributed. The main challenge  in this asymptotic analysis is
to quantify the order of  an approximation  of the quantity 
\begin{equation}  \label{massest}
\lambda \big ( 
    \{ t \in [0,1] \colon   | \hat \mu (t) -  \hat \mu ({0}) | > c \}   \big) ,
\end{equation}
where $\hat \mu$ is an  appropriate estimate of the regression function. While estimates  of the mean trend have been already  studied  under local stationarity in the literature [see, for example, \cite{wu2007inference}],  the analysis of the quantity \eqref{massest}  and its approximation requires a 
careful localization of the effect of the estimation error around the critical roots satisfying the equation $|\mu(t) - \mu(0) |=c $.

It is  demonstrated -  even in the case of independent or stationary  errors -  that the
  variance of the limit distribution depends sensitively  on (eventually higher order) derivatives of the regression function at the critical roots, which 
   are very difficult to estimate. 
Moreover, because of the  non-stationarity of the error process in \eqref{protype}  the asymptotic variance depends also  in a complicated way on the unknown dependence structure.
 We propose a bootstrap method to obtain critical values for the test, which is motivated by a Gaussian approximation  used
 in the proof of the  asymptotic normality.  This re-sampling procedure  is adaptive in the sense that it avoids the direct estimation of the critical roots and
the values of the derivatives of the regression function 
at these points.

 Note that  $T_c$ is the {\it excess} Lebesgue measure (or mass) of the time when the absolute difference between   the mean trend
 and its initial value exceeds the  level $c$.  Thus our  approach is    naturally related  to the concept of  excess  mass
 which has found considerable attention in the literature. Many authors used the excess  mass  approach   to investigate  multimodality of a density  [see, for example, \cite{muesaw1991},  \cite{polonik1995}, \cite{chehall1998}, \cite{polowang2005}].
The  asymptotic properties of  distances  between an estimated level and the ``true'' level set of a density
have also been studied in several publications  [see  \cite{baillo2003}, \cite{cadre2006}, \cite{cuevasetal2006} and \cite{mason2009}
 among many others].  The concept of mass excess  has additionally been used for  discrimination   between time series [see \cite{chanpolo2006}], for the construction of monotone regression estimates  [\cite{DNP2006}, \cite{CFG2010}],  quantile regression [\cite{detvol2008}, \cite{cher2009}], clustering [\cite{rinaldo2010generalized}]
  and for bandwidth selection in density estimation [see \cite{samworth2010}], but to our best knowledge it has not been used for change point analysis.

Most of the literature  discusses  regular points,
  that  {are}   points, where the first derivative of the density or regression function does not vanish, but 
there exist  also  references where this condition is relaxed.
 For example, \cite{hartigan1985dip}  proposed a test  for  multimodality of a density comparing the difference between the empirical distribution function and a class of unimodal distribution functions.  They observed that the stochastic order of the test statistic depends on the minimal number $k$, such  that
the $k${th} derivative of the cumulative distribution function does not vanish.
\cite{polonik1995} studied the asymptotic properties  of an estimate of  the   mass excess
functional  
of a cumulative distribution function $F$ with density $f$
and
\cite{tsybakov1997nonparametric}
observed that the minimax risk in the problem of estimating the level set of a density depends on its
``regularity''. More recently,
\cite{chanpolo2006} used the excess mass functional for discrimination analysis under the additional assumption of  unimodality.

The present paper  differs from this literature with respect to several perspectives. First, we are interested in change point analysis and develop a test for a relevant difference in the mean of the process over a certain range of time.  Therefore - in contrast to most of the literature, which deals with i.i.d. data -   we consider the regression model \eqref{protype} with a non-stationary error process.  Second, we are interested in an estimate, say $\hat T_{N,c}$ of the Lebesgue measure $ T_{c} $ of the level set ${\cal M}_c$ and its  asymptotic properties   in order to construct a test for the change point problem \eqref{nullnew}. Therefore - in contrast to many references -  we do not discuss  estimates of  an excess mass functional or   a distance between  an estimated level set and the ``true'' level set, but 
investigate the  asymptotic  distribution of $\hat T_{N,c}$.
Third,  as this distribution depends sensitively on the critical points and the dependence structure of the non-stationary error process,
we  use a Gaussian approximation to develop a
bootstrap method, which  allows us to  find quantiles without estimating  the  location of the critical points and the derivatives of 
the regression function at these points. 

We also mention the differences to the work of 
\cite{mercurio2004} and \cite{spokoiny2009}, which has its focus on the detection of intervals of homogeneity of the underlying process, while the present paper investigates the problem to detect  significant deviations
of an inhomogeneous process from its  initial distribution (here specified by different values of the mean function).

{The approach proposed in this paper is  also related  to the  sojourn time of a (real valued) stochastic process, say $\{X(t) \}_{t \in [0,1]}$, which is defined as
\begin{eqnarray} 
\label{soj11}
S_c = \int_0^1 \mathbf 1 \{ |X (t)- X (0) | > c\}  dt 
\end{eqnarray}
and has widely been studied in probability theory under specific distributional assumptions [see, for example \cite{berman1992,takacs1996} among many]. To be precise let  $ X(t) = \mu (t) + \epsilon (t) $ for some centered process  $\{\epsilon (t) \}_{t \in [0,1]}$, then compared to the quantity  $T_c$ defined in \eqref{tc}, which refers to expectation $\mu(t)$, the quantity  $S_c$ is a random variable. An alternative excess-type measure is now given by  the expected sojourn time
\begin{align} \label{soj21}
e_c:= \E(S_c)  ~,
\end{align}
and the corresponding null hypotheses can be formulated as 
$$H_0:  e_c \leq \Delta \mbox{~~versus ~~} H_1: e_c >\Delta.
$$
A further quantity of interest was mentioned by a referee to us and 
is defined by the probability that the sojourn time exceeds the threshold $\Delta$, that is
\begin{align} 
\label{soj22}
p_{c,\Delta}:= \mathbb{P} (S_c > \Delta)  .
\end{align}
This quantity cannot be directly used for testing, but can be considered as a measure of a relevant deviation for a sufficiently long time from the initial state
$X(0)$. 
         }
         
The rest of paper is organized as follows. In Section \ref{sec2}  we motivate our approach, 
define an estimator of the quantity
$T_c$, discuss alternative measures and
give some basic assumptions of the non-stationary model \eqref{protype}. Section \ref{sec3} is devoted
to a discussion of the asymptotic properties of this estimator in the case, where all critical points are regular points, that is $\mu^{(1)} (s)   \not =0 $ for all $ {s} \in {\cal C}$. We focus on this case first, because here the arguments are  more transparent. 
 In particular in this case all roots are of the same order
and  contribute to the asymptotic variance of the limit distribution, which simplifies the statement of the results substantially. 
In this case we also identify a bias problem, which makes the implementation
of the test at this stage difficult.
The general case is carefully investigated in Section \ref{sec4}, where we also address the bias problem using a Jackknife approach.  The bootstrap procedure is developed
in the second part of Section \ref{sec4}. In Sections \ref{sec6} and \ref{sec62} we illustrate its finite sample properties by means of a simulation study and by analyzing data examples. 
Finally, some discussion on multivariate data is 
 given in Section \ref{sec8}. In this section we also  propose estimators of the quantities \eqref{soj21} and \eqref{soj22}.
Finally, most of the technical  details are deferred to  Section \ref{sec7} and 
an online supplement (which also contains some further auxiliary results).

\section{Estimation and basic assumptions} \label{sec2}
\def\theequation{2.\arabic{equation}}
\setcounter{equation}{0}

\subsection{Relevant changes via a mass excess approach} \label{sec21}

Recall the  definition of the testing problems \eqref{nullnew}, \eqref{nullnew+}, \eqref{nullnew-} and note that
$T_c = T_c^+ + T_c^-$, where
\begin{equation}
\label{tc+}
T_c^+= \int_0^1 \mathbf 1 (\mu(t)-\mu(0)> c)dt ~,~~ 
T_c^{-}= \int_0^1 \mathbf 1 (\mu(t)-\mu(0)< -c)dt,
\end{equation}
and $\mathbf 1 (B) $ denotes the indicator function of the set $B$. In most parts of the paper we mainly
concentrate on the estimation of the quantity $T_c^+$
and study the asymptotic properties of an appropriately standardized estimate [see for example Theorems \ref{Thm1} and
\ref{criticalversion}].
Corresponding results for the  estimators of  $T_c^-$  and   $T_c$ can be obtained by similar methods and the
joint weak convergence is established in  Theorem \ref{Thm2} and Theorem \ref{Thm4} without giving detailed proofs.

We propose to estimate  the mean function by a local linear
estimator 
\begin{align}\label{Locallinear}
(\hat \mu_{b_n}(t), \hat{ \dot {\mu}}_{b_n}(t))^T=\argmin_{\beta_0\in \mathbb R,\beta_1\in \mathbb R}\sum_{i=1}^n\left(X_i-\beta_0-\beta_1(i/n-t)\right)^2K\Big(\frac{i/n-t}{b_n}\Big)~,
t\in[0,1] \end{align}
where $K(\cdot)$ denotes a continuous and symmetric kernel supported on  the interval $[-1,1]$. 
We define  an estimator of $T_c^+$ by
\begin{align} \label{stat}
\hat{T}^+_{N,c}=\frac{1}{N}\sum_{i=1}^N\int_c^{\infty}\frac{1}{h_d}K_d\Big (\frac{\hat{\mu}_{b_n}(i/N)-\hat{\mu}_{b_n}(0)-u}{h_d}\Big  )du,
\end{align}
 where $K_d(\cdot)$ is a symmetric kernel function supported on the  interval  $[-1,1]$
 such that $\int_{-1}^1 K_d(x)dx=1$.  In \eqref{stat}
 the quantity
 $h_d>0$ denotes a bandwidth and  $N$ is the number of knots in a  Riemann approximation (see the discussion in the following paragraph), which  does not need to coincide with the sample size  $n$. It turns out that the procedures  proposed in this paper are not sensitive with respect to the choice of $h_d$ and $N$, provided that these parameters have been chosen sufficiently small  and large, respectively (see Section \ref{sec6} for a further discussion). 
 
\begin{figure}[htbp]
	\centering
		\includegraphics[width=7cm,height=5cm]{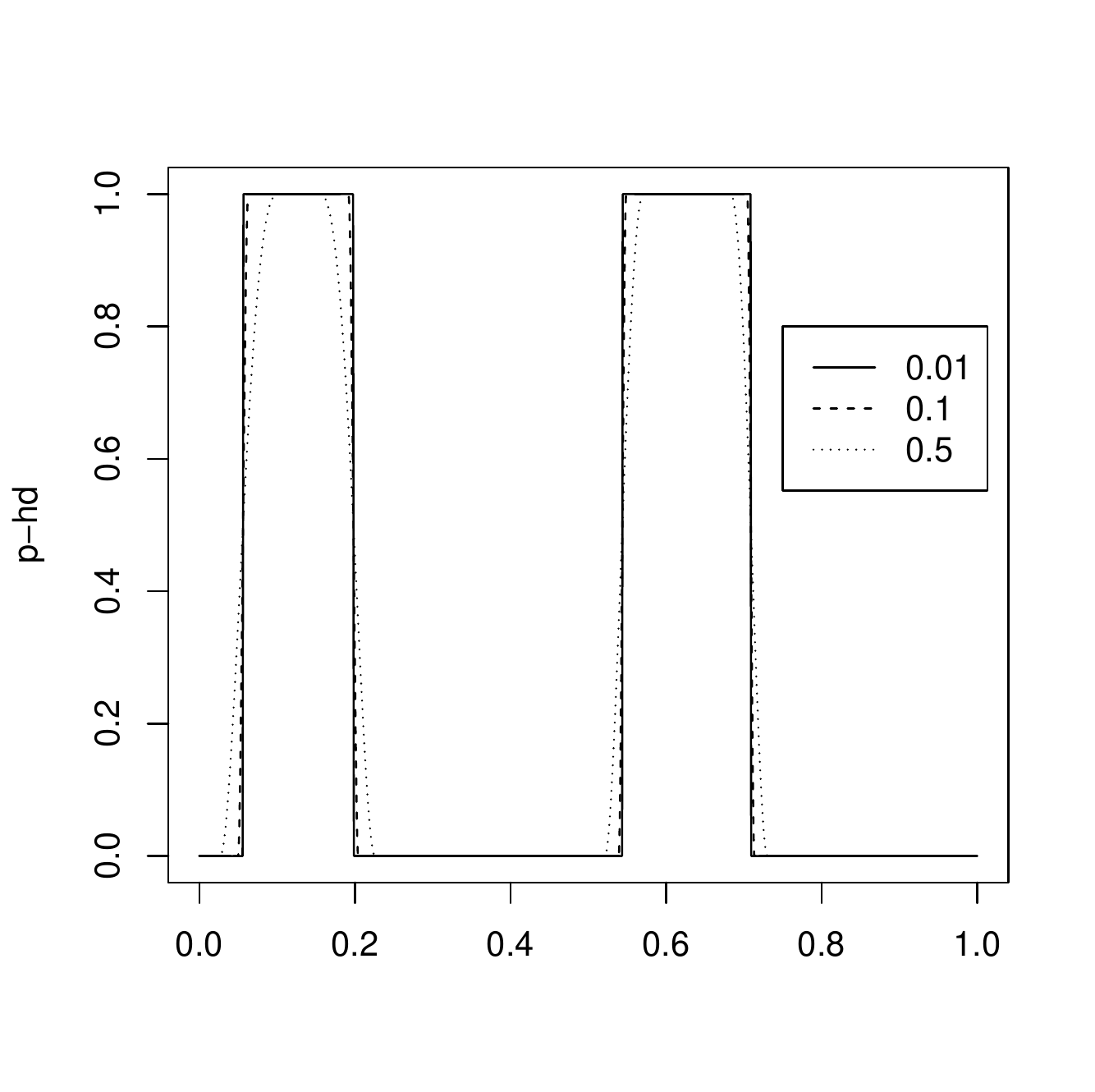}
		\caption{ \it Smooth approximation $p_{h_d}$ of the step function  $q = \mathbf{1}_{\mathcal {M}_c^+}$ for different choices of the bandwidth $h_d$. \label{illufirst}}
\end{figure}

A statistic  of the type \eqref{stat} has been proposed by \cite{DNP2006} to estimate the inverse of a strictly increasing regression function,
but we use it here without assuming monotonicity of the mean function $\mu$.
Observing that $\hat \mu_{b_n}(t)$ is a consistent estimate of $\mu (t)$ we argue
(rigorous  arguments are given later) that
\begin{align}
\hat{T}^+_{N,c} &=  \frac{1}{N}\sum_{i=1}^N\int_c^{\infty}\frac{1}{h_d}K_d\Big  (\frac{{\mu}(i/N)-{\mu}(0)-u}{h_d}\Big  )du + o_P(1) \nonumber\\ %
&= \frac{1}{h_d} \int^1_0 \int_c^{ \infty} K_d \Bigl( \frac{\mu(x) - \mu(0) - u}{h_d} \Bigr) dudx + o_P(1)  
= T_c^+  + o_P(1)   \label{lem1}
\end{align}
as $n, N \to \infty$, $h_d \to 0$. In Figure \ref{illufirst} we display
the functions
\begin{align} \label{illu0}
p_{h_d}: t \to
\frac{1}{h_d} \int_c^{ \infty} K_d \Bigl( \frac{\mu(t) - \mu(0) - u}{h_d} \Bigr) du
\mbox{ ~~ and ~~} q: t \to  \mathbf 1(\mu(t)-\mu(0)\geq c)
\end{align}
and visualize  that $p_{h_d}$ is a smooth approximation of the indicator function for decreasing $h_d$ (for the function considered in  Figure \ref{illustration}). This smoothing is introduced to derive the asymptotic properties of the statistic  $\hat{T}^+_{N,c}$ and to construct a valid bootstrap procedure 
without estimating the critical roots and  derivatives of the regression function.  
Thus intuitively (rigorous arguments will be given in the following sections) the statistic $\hat{T}^+_{N,c}$ is a consistent estimator of $T_c^+$ and a similar argument for $T_c^-$ will provide a 
consistent estimator of the quantity $T_c$ defined in \eqref{tc}.
The null hypothesis is finally rejected for large values of this estimate.

In order to make these heuristic arguments more rigorous we  make the following basic assumptions for the  model \eqref{protype}.
\begin{assumption} \label{A1} {\rm ~~
\begin{description}
\item[(a)] The mean function is twice  differentiable with Lipschitz continuous second derivative.
\item[(b)]
There exists a positive constant $\epsilon_0$, such that for all $\delta \in [0,\epsilon_0]$
there are $k_\delta$  closed disjoint intervals $I_{1,\delta},\ldots ,I_{k_\delta,\delta}$,
 such that
\begin{align*}
\big \{t\in[0,1]:|\mu(t)-\mu(0)-c|\leq \delta \big\}\cup\big \{t\in[0,1]:|\mu(t)-\mu(0)+c|\leq \delta\big \}=\bigcup_{i=1}^{k_\delta} I_{i,\delta}, 
 \end{align*}
where the  number of intervals $k_\delta$ satisfies 
{$\sup_{0 \leq  \delta \leq \epsilon_0}k_{\delta}\leq  M$
for some universal constant $M$.} In particular there exists only a finite number of roots of the equation
$\mu (t) - \mu (0) = \pm c$. We also assume that $ | \mu (1) - \mu (0) | \not= c$.
\end{description}
}
\end{assumption}
\smallskip

It is worthwhile to mention that all results presented in the paper remain true
if the regression function is Lipschitz continuous on the interval $[0,1]$ and 
the assumptions regarding its differentiability (such as Assumption \ref{A1}) hold
in a neighborhood of the critical  roots. 
Our first result makes the approximation of $T_c^+$ by
its deterministic counterpart 
\begin{align} \label{tdet}
{T}^+_{N,c}:=\frac{1}{N}\sum_{i=1}^N\int_{c}^\infty
\frac{1}{h_d}K_d\Big (\frac{{\mu}(i/N)-{\mu(0)}-u}{h_d}\Big )du
\end{align}
in \eqref{lem1} rigorous. For this purpose let
\begin{align} 
\label{mcd}
m_{\gamma,\delta}(\mu)=\lambda \big 
( \{ t\in [0,1]: |\mu(t)-\gamma |\leq \delta \} \big )
\end{align}
denote the Lebesgue measure of the set of points, where the mean function  lies in a $\delta$-neighbourhood of the point  $\gamma$.
\begin{proposition}\label{prop1}
If Assumption  \ref{A1}  holds and  $m_{c+\mu(0),\delta}(\mu)=O(\delta^\iota)$ for some $\iota>0$ as $\delta\rightarrow 0$, we have for the quantity $T^+_{N,c}$ in \eqref{tdet},
$$
T^+_{N,c}-T^+_{ c}=O(\max\{h_d^\iota,N^{-1}\})
$$
as $N  \to \infty$,  $h_d\rightarrow 0$.
\end{proposition}
\noindent
{\it Proof.}  By elementary calculations it follows that
\begin{align*}
&\int_{c}^\infty\frac{1}{h_d}K_d\Big(\frac{{\mu}(i/N)-{\mu(0)}-u}{h_d}\Big)du-\mathbf 1(\mu(i/N)-\mu(0)> c)\notag\\
&=\mathbf 1( \{|c-(\mu(i/N)-\mu(0))|\leq h_d \} )\int_{\frac{c-\mu(i/N)+\mu(0)}{h_d}}^{\infty}K_d(x)dx\notag \\&-\mathbf 1(
\{\mu(i/N)-\mu(0)-h_d< c\leq\mu(i/N)-\mu(0) \} ).
\end{align*}
Therefore, we obtain  (observing that
$\int_{-1}^1 K_d(x)dx=1$)
\begin{align*}
| T^+_{N,c}-T^+_{c} | &=
 \left| \frac{1}{N}\sum_{i=1}^N \left(\int_{c}^\infty\frac{1}{h_d}K_d\Big (\frac{{\mu}({i\over N})-{\mu(0)}-u}{h_d}\Big )du -  \mathbf 1(\mu(\tfrac{i}{N})-\mu(0)> c)\right)
 \right|  + O\Big({1 \over N}\Big )\\
&  \leq   \frac{2}{N}\sum_{i=1}^N\mathbf 1( |\mu(i/N)-\mu(0)-c|\leq h_d  )  + O(N^{-1}) \\
& =2  m_{c+\mu(0),h_d}(\mu) + O(N^{-1}) =O\Big(\max\{h_d^\iota,{1 \over N}  \}\Big) .
\end{align*}
as $N  \to \infty$,  $h_d \to 0$.
 \hfill $\Box$

\subsection{Alternatives measures of mass excess} \label{sec22}
In this section we briefly mention several alternative measures
of  mass excess, which might 
be of interest in applications and for which similar results as stated in this paper
can be derived. For the sake of brevity we do not state
these results in full detail in this paper and only describe the measures with corresponding estimates.

\subsubsection{Deviations from an average trend} \label{sec221}
In applications one might be 
also interested if there exist relevant deviations of the sequence $(\mu (i/n) )_{i= \lfloor n t_0 \rfloor +1,\ldots , n} $ from an average trend formed from  the previous period $(\mu (i/n) )_{i=1,\ldots , \lfloor n t_0 \rfloor} $. 
This question can be addressed by estimating  the  quantity 
$$
\int_{t_0}^1 \mathbf 1\Big (\mu(t)-\int_{0}^{t_0}\mu(s)ds > c \Big )dt = \lambda  \Big ( \big\{ 
t \in [t_0,1] \colon ~ \mu(t)-\int_{0}^{t_0}\mu(s)ds > c
\big \} \Big ).
$$
Using similar arguments as given in this paper (and the supplementary material) one can prove consistency and derive the asymptotic distribution of the  estimate
$$
\frac{1}{N}\sum_{i=\lf Nt_0\rf}^N\int_{c}^\infty\frac{1}{h_d}K_d\Big (\frac{\hat \mu_{b_n}(i/N)-\int_0^{t_0}\hat \mu_{b_n}(s)ds-u}{h_d}\Big )du
$$
where $\hat \mu_{b_n} $ is  local linear estimator of $\mu$ (in Section \ref{sec4} we will use a bias corrected version of $\hat \mu_{b_n} $). 
\subsubsection{Relative deviations} \label{sec222}
If $\mu (0) \not =0 $	an alternative measure of excess can be defined
by
\begin{align}
\int_{0}^1 \mathbf 1\Big (\Big | \tfrac{\mu(t)- \mu(0 )}{\mu(0 )} \Big |  > c \Big )dt = \lambda  \Big ( \Big\{ 
t \in [0,1] \colon ~ \Big | \tfrac{\mu(t)- \mu(0 )}{\mu(0 )} \Big | > c  
\Big \} \Big ).\label{relmass}
\end{align}
This measure of excess allows to define a relevant  change in the mean relative to its initial value
and makes the choice of the constant $c$ easier in applications. 
For example, if one chooses $c=0.1$, one is interested in relevant deviation from the initial value by more than 10\%. The quantity in equation \eqref{relmass}  can be estimated in a similar way as described in the previous paragraph and the details are omitted for the sake of brevity.

\subsection{Locally stationary processes}
\label{sec23}

 In Sections \ref{sec3} and \ref{sec4}  we will establish the asymptotic properties of  the statistic $\hat{T}^+_{N,c}$ as an estimator of $T_c^+$ and derive a bootstrap approximation to derive critical values. Since we are interested in a procedure for non-stationary processes we require several technical assumptions on the error process in model \eqref{protype}. The less experienced reader can easily skip this paragraph  and consider an independent identically distributed array of centered random variables  $\epsilon_{i,n}$ in model \eqref{protype} with variance $\sigma^2$. The main challenge in the proofs is neither the dependence structure nor the non-stationarity of the error process but  
consists in the fact that definition \eqref{stat} defines a complicated map from the class of estimators to the Lebesgue measure of  random sets of the form  $\{  t:|~\hat\mu_{b_n} (t) - \hat\mu_{b_n} (0) | > c \}$. Thus,  although  a standardized version of the local linear estimator $\hat\mu_{b_n}$ is asymptotically normally distributed
(under suitable conditions),  a rigorous analysis of this mapping is required to derive  the  distributional properties of the statistic $\hat{T}^+_{N,c}$. 
These depend sensitively on the local behaviour of the function $\mu $ at points satisfying the equation $| \mu (t) - \mu (0) | =c$  and the  corresponding analysis  represents the most important part of the work, which is independent of the error structure in model \eqref{protype}.

To be precise let  $|| X ||_q  = \big ( \E |X|^q  \big )^{1/q}$ denote the $\mathcal L_q$-norm of the random variable $X$ ($q \geq 1$).
We begin recalling some basic definitions  on physical dependence measures and locally stationary processes.

\medskip

\begin{definition}
Let  $\eta = (\eta_i)_{i\in \mathbb Z}$  be a sequence of independent identically distributed  random variables, $\FF_i= \{\eta_s \colon s\leq i \} $,
denote by  $\eta' = (\eta_i')_{i\in \mathbb Z}$ an independent  copy of $\eta$ and define
$\FF_i^*=(\ldots ,\eta_{-2},\eta_{-1},\eta_0',\eta_1,\ldots,\eta_i)$. For $t \in [0,1]$  let $G: [0,1] \times  \R^\infty  \to \R $ denote a nonlinear
filter, that is  a measurable function, such that  $G(t, \FF_i)$ is a properly defined random variable for all $t\in [0,1]$.
\begin{description}
\item[(1)] A sequence $(\epsilon_{i,n} )_{i=1,\ldots , n}$ is called locally stationary process, if there exists
a filter  $G$  such  that $\epsilon_{i,n} = G(i/n, \FF_i)$ for all $i=1,\ldots , n$.
\item[(2)] For a  nonlinear filter
$G$ with  $\sup_{t\in [0,1]}\|G(t,\FF_i)\|_q<\infty$, the physical dependence measure of $G$ with respect to $\|\cdot\|_q$
is defined by
\begin{align}
\delta_q(G,k)=\sup_{t\in [0,1]}\|G(t,\FF_k)-G(t,\FF_k^*)\|_q.\label{new.2.10}
\end{align}
\item[(3)] The filter $G$  is called Lipschitz continuous with respect to $\| \cdot  \|_q$
 if and only if
  \begin{align}\sup_{0\leq s<t\leq 1}\|G(t,\FF_i)-G(s,\FF_i)\|_q/|t-s|<\infty.\label{new.2.11}
\end{align}
\end{description}
\end{definition}
\medskip
\noindent The filter $G$ is used to model non-stationarity. The quantity $\delta_{q}(G,k)$ measures the dependence
of $G(t,\FF_k)$ on $\eta_0^\prime$ over the interval $[0,1]$. When $\delta_{q}(G,k)$ converges sufficiently fast to $0$  such that $\sum_k\delta_{q}(G,k)<\infty$, we
speak of a short range dependent time series. Condition \eqref{new.2.11} means that the data generating mechanism $G$ is varying smoothly  in time.  We refer to \cite{zhou2009local} for more details, in particular for examples of locally stationary linear and nonlinear time series, calculations of the dependence measure \eqref{new.2.10} and for the verification of \eqref{new.2.11}.  With this notation we make the following assumptions regarding the error process in model \eqref{protype}.

\begin{assumption} \label{A2} {\rm
The error process $(\epsilon_{i,n} )_{i=1,\ldots , n}$
in model \eqref{protype}  is a zero-mean locally stationary process with filter $G$, which satisfies the following conditions:
\begin{description}
\item[(a)] There exists a constant $\chi\in (0,1)$, such that {
$\delta_4 (G,k)=O(\chi^k)$  as $ k \to \infty$.}
\item[(b)]  The filter $G $  is Lipschitz continuous with respect to   $\| \cdot  \|_4$ and $\sup_{t\in[0,1]}\|G(t,\FF_0)\|_4<\infty$.
\item[(c)] The {\it long-run variance}  
\begin{align} \label{longrun}
\sigma(t):=\sum_{ {i}=-\infty}^\infty\mbox{cov} (G(t,\FF_i),G(t,\FF_0)),~~ t\in [0,1]
\end{align}
of the filter $G$ is Lipschitz continuous on the interval  $[0,1]$ and  {\it  non-degenerate}, that is $\inf_{t\in[0,1]}\sigma(t)>0$.
\end{description}
}
\end{assumption}

\noindent
 Condition (a) of Assumption \ref{A2} means that the error process $\{\epsilon_{i,n}\}_{i=1,...,n}$
in model \eqref{protype} is locally stationary with geometrically decaying dependence measure. The theoretical results of the paper can also be  derived under the assumption of a polynomially decaying dependence measure with  substantially more  complicated bandwidth conditions and proofs. Conditions (b) and (c) are standard in the literature of locally stationary time series. They
are used later for a   Gaussian approximation of the locally stationary time series; see for example   \cite{zhou2010simultaneous}.

\section{Twice continuously differentiable mean functions} \label{sec3}
\def\theequation{3.\arabic{equation}}
\setcounter{equation}{0}

In this section we briefly consider the situation, where the derivatives  of the mean function at the critical set $\mathcal C$ do not vanish. These  assumptions are quite common in the  literature  [see, for example, condition (B.ii) in \cite{mason2009} or assumption (A1) in \cite{samworth2010}]. We discuss this case separately  because of (at least) two reasons.   First, the results and required assumptions are slightly simpler here. Second, and more important, we use this case to demonstrate  that  the estimates of $T_{c}$, $T_{c}^+$ and  $T_{c}^-$ have a bias, which is asymptotically not negligible and  makes their direct application for testing the hypotheses \eqref{nullnew}, 
\eqref{nullnew+} and \eqref{nullnew-} difficult.
 The general case is postponed to Section \ref{sec4}, where we solve the bias problem and also introduce a bootstrap procedure. We do not provide proofs of the results in this section, as they can be obtained by similar (but substantially simpler) arguments as given in the proofs of  Theorems \ref{criticalversion} and \ref{Thm4} below. 
 
Recall the definition of the statistic $\hat T_{N,c}^+$ in \eqref{stat}, where $\hat \mu_{b_n}(t)$ is the local linear estimate of the mean
function with bandwidth $b_n$. Our first result specifies its asymptotic distribution, and for its statement we make the following additional assumption on the  bandwidths. 

\begin{assumption}\label{A3} 
 The bandwidth $b_n$ of the local linear estimator satisfies $b_n\rightarrow 0, nb_n\rightarrow \infty$, $b_n/h_d\rightarrow \infty,$ $\sqrt{n}b_n/\log^4 n\rightarrow \infty,$ and $\pi^*_n/h_d\rightarrow 0$ where
$$
\pi^*_n:=(b_n^2+(nb_n)^{-1/2}\log n)\log n.
$$
\end{assumption}

\begin{theorem}\label{Thm1}
Suppose that  Assumptions \ref{A1},  \ref{A2} and  \ref{A3} hold, that there exist  roots 
$t_1^+, \ldots , t_{k^+}^+$ of the equation $\mu(t)-\mu(0)=c$ satisfying
$\dot{\mu}({t_j^+})\neq 0$ for $1\leq j\leq k^+$,
and  define
\begin{align*} \bar R_{1,n}
& =\frac{n^{1/4}\log^2n}{nb_n}, ~\bar R_{2,n}=\Big (\frac{1}{Nb_n}+\frac{1}{Nh_d} \Big )(b_n\wedge h_d), \\
\bar \chi_n & =(b_n^4+\frac{1}{nb_n})h_d^{-1}.
\end{align*}
 If  $Nb_n\rightarrow \infty$, $Nh_d\rightarrow \infty$, $\sqrt{nb_n}(\bar \chi_n+\bar R_{1,n}+\bar R_{2,n})=o(1),$
then
\begin{align*}
\sqrt{nb_n}\Big (\hat T^+_{N,c}-T_{c}^+ -\mu_{2,K}b_n^2\sum_{j=1}^{k^+}\frac{\ddot \mu(t^+_j)}{|\dot \mu(t^+_j)|}+\frac{b_n^2c_{2,K}\ddot \mu(0)}{2c_{0,K}}\sum_{j=1}^{k^+}\frac{1}{|\dot \mu(t^+_j)|}\Big )
\convd
{\cal N} (0,\tau_1^{2,+} + \tau_2^{2,+}),
\end{align*}
{
where
\begin{align*}
\tau_1^{2,+} &=\sum_{s=1}^{k^+}\frac{\sigma^2(t_s^+)}{\dot \mu(t_s^+)^2}\int K^2(x)dx, \\
\tau_2^{2,+} &={\sigma^2(0) \over c_{0,K}^{2}} \Big (\sum_{j=1}^{k^+}\frac{1}{|\dot \mu(t_j^+)|}\Big)^2\int_0^1\big (\mu_{2,K}-t\mu_{1,K} \big )^2K^2(t)dt,
\end{align*}
 the  constants  $c_{0,K}$ and   $c_{2,K}$  are given by
 \begin{align*}
&c_{0,K}=\mu_{0,K}\mu_{2,K}- \mu_{1,K}^2~, ~~ c_{2,K}= \mu_{2,K}^2-\mu_{1,K}\mu_{3,K}
\end{align*}
and $
\mu_{l,K}=\int_0^1x^lK(x)dx$ for ($l =1,2,\ldots $).
}
\end{theorem}

\medskip
\noindent
 Theorem \ref{Thm1} establishes  asymptotic normality under the scenario that ${\dot{\mu}}(t)\neq 0$ for  all points $t\in
 {\cal C}^+ =\{t\in[0,1] \colon \mu(t)-\mu(0)=c\}$. This condition guarantees that
 the mean function $\mu$ is strictly monotone in a neighbourhood of the roots.
Moreover, \ref{A1}(b), Assumption \ref{A2} and  \ref{A3}
 imply the asymptotic independence of the estimators of $\mu (0)$ and $\mu (t) $ for any $t \in  {\cal C}^+$.
 
 We conclude this section presenting a corresponding weak convergence result  for the joint distribution of $(\hat T^+_{N,c}, \hat T^-_{N,c})$,
 where
 \begin{align} \label{statmin}
 \hat{T}^-_{N,c}=\frac{1}{N}\sum_{i=1}^N\int_{-\infty}^{-c}\frac{1}{h_d}K_d\left(\frac{\hat{\mu}_{b_n}(i/N)-\hat{\mu}_{b_n}(0)-u}{h_d}\right)du
 \end{align}
 denotes an estimate of the quantity $T^-_c$ defined in \eqref{nullnew-}.

\medskip

 \begin{theorem}\label{Thm2}
Suppose that  Assumptions \ref{A1},  \ref{A2} and  \ref{A3} are satisfied and that the {bandwidth conditions of Theorem \ref{Thm1} hold. If there also exist roots  
 $t_1^-, \ldots , t_{k^-}^-$   of the equation $\mu(t)-\mu(0)=-c$, such that 
 $\dot \mu(t^-_j) \not =0$  $(j=1,\ldots , k^-)$, 
then, as $n \to \infty$,
 \begin{align}
\sqrt{nb_n} \left(
\hat T^+_{N,c}-T_{c}^+ -\beta^+_c , 
\hat T^-_{N,c}-T_{c}^-  - \beta^-_c
\right)^T
\convd  {\cal N} (0,\tilde \Sigma),
 \end{align}
 where
 \begin{align} \label{bias}
\beta^\pm _c ={\mu_{2,K}b_n^2}\sum_{j=1}^{k^\pm }\frac{\ddot \mu(t^\pm _j)}{|\dot \mu(t_j^\pm )|}-\frac{b_n^2c_{2,K}\ddot \mu(0)}{2c_{0,K}}\sum_{j=1}^{k^\pm }\frac{1}{|\dot \mu(t^\pm _j)|},
 \end{align}
 and  the elements in the matrix $\tilde \Sigma = ( \tilde \Sigma_{ij} )_{i,j=1,2} $ are given by
 $ \tilde \Sigma_{11} =\tau_1^{2,+}  + \tau_2^{2,+}$, $\tilde \Sigma_{22}=\tau_1^{2,-}  + \tau_2^{2,-}  $ and  
 \begin{align*}
 \tilde \Sigma_{12}& =\tilde \Sigma_{21}=-c_{0,K}^{-2}\sigma^2(0)\Big (\sum_{j=1}^{k^+}\frac{1}{|\dot \mu(t_j^+)|}\Big )\Big (\sum_{j=1}^{k^-}\frac{1}{|\dot \mu(t_j^-)|}\Big) \int_0^1(\mu_{2,K}-t\mu_{1,K})^2K^2(t)dt .
 \end{align*}
 where $\tau_1^{2,-} $ and $   \tau_2^{2,-}$   are defined in a similar way as $\tau_1^{2,+}$  and $ \tau_2^{2,+} $ in Theorem \ref{Thm1}.}
\end{theorem}

\medskip

\begin{remark} 
{\rm The representation of the bias in \eqref{bias} has some similarity with the approximation of the risk of an estimate of the highest density region 
investigated  in \cite{samworth2010}. We suppose that similar arguments as given in the proofs of our main results can be used to derive asymptotic normality of this estimate [see also \cite{mason2009}].}
\end{remark}
\noindent
\begin{remark} \label{generalassumption}

{\rm  The most general assumptions under which the  results of  our paper hold are the following.
    \begin{description}
    	\item (a) The mean trend  is a piece-wise Lipschitz continuous function, with a bounded number of jump points. 
{     If $D^+ (t_0)$ and $D^- (t_0)$ denote the  limit of the function $|\mu (\cdot  ) - \mu (0)|$ from the left an right at the jump point $t_0$, then
  $ (D^+ (t_0) -c)(D^- (t_0) -c) >0$. In other words: at any jump, the function  $|\mu (\cdot  ) - \mu (0)|$ does not ``cross'' the level  $c$.}
    	\item (b) There is a finite number of critical roots and
        the mean trend function has a Lipschitz continuous second derivative in a neighborhood of each critical root. 
    \end{description}
    In particular we exclude the case where jumps occur at critical roots, but there might be jumps at other points in the interval $[0,1]$. 
    In this case the local linear estimator $\hat \mu_{b_n}$ has to be modified to address for these jumps [see \cite{qiu2003jump} or  \cite{Gijbels2007} among others]. For the sake of a transparent representation and for the sake of brevity we state our results under Assumption \ref{A1} and \ref{A2}. 
  }  
  
\end{remark}

Theorem \ref{Thm1} and \ref{Thm2} can be used to construct tests for the hypotheses \eqref{nullnew+} and  \eqref{nullnew-}.
Similarly, by  the continuous mapping theorem we also obtain from Theorem \ref{Thm2} the asymptotic distribution of the the statistic $\hat T_{N,c} = \hat T_{N,c}^+ + \hat T_{N,c}^- $, which could  be used to construct a test for the hypotheses \eqref{nullnew}. However, such  tests would either require undersmoothing or estimation of the bias
$\beta_c^+ $ and $\beta_c^-$ in \eqref{bias}, which is not an easy task.  
We address this problem by a Jackknife method in the following section where  we also develop a bootstrap test to avoid the estimation of the critical roots.

\section{Bias correction and bootstrap} \label{sec4}
\def\theequation{4.\arabic{equation}}
\setcounter{equation}{0}

In this section we will address the bias problem mentioned in the previous section  adopting the Jackknife bias reduction technique proposed by \cite{schucany1977}. In a second step we will use these results to construct a bootstrap procedure. Moreover, we also  relax the main assumption 
in Section \ref{sec3} that the derivative of the mean  function does not vanish at  critical roots $t \in {\cal C}$.

\subsection{Bias correction}
Recalling the definition $\hat \mu_{b_n}(t)$ of the local linear estimator  in  \eqref{Locallinear} with bandwidth $b_n$
we define the Jackknife estimator by  
\begin{align}\label{Jack}
\tilde \mu_{b_n}(t)=2\hat \mu_{b_n/\sqrt 2}(t)-\hat \mu_{b_n}(t)
\end{align}
for $0\leq t\leq 1$. 
It has been shown in \cite{wu2007inference} 
that the  bias of the estimator \eqref{Jack} is of order  $o(b_n^3+\frac{1}{nb_n})$, whenever  $b_n\leq t\leq 1-b_n$,
and \cite{zhou2010simultaneous} showed  that the estimate $\tilde \mu_{b_n}$
is asymptotically equivalent
to a local linear estimate with kernel
\begin{align} \label{k1}
K^*(x)=2\sqrt 2K(\sqrt 2 x)-K(x).
\end{align}
In order to use these bias corrected estimators for the  construction of   tests  for the hypotheses defined in \eqref{nullnew} - \eqref{nullnew-}, {we also need to study the estimate  $\tilde \mu_{b_n}(0)$, which
is not asymptotically equivalent to a local linear estimate with kernel
$K^*(x)$. However, as  a consequence of  Lemma \ref{Jackknife Approximation} in the online supplement we obtain the stochastic expansion
\begin{align}
\Big |\tilde \mu_{b_n}(0)-\mu(0)-\frac{1}{nb_n}\sum_{i=1}^n\bar K^*(\frac{i}{nb_n})\epsilon_{i,n}\Big |=O(b_n^3+\frac{1}{nb_n}),
\end{align}
where the kernel $\bar K^*(x)$ is given by  }
\begin{align} \label{k2}
\bar K^*(x)=2\sqrt 2\bar K(\sqrt 2 x)-\bar K(x)\end{align}
 with
$ \bar K(x)=({\mu_{2,K}-x\mu_{1,K}})K(x)/{c_{0,K}}.$
Since the kernel $\bar K^*(x)$ is not symmetric, the bias of $\tilde \mu_{b_n}(0)$ is of the order $O(b_n^{3}+\frac{1}{nb_n})$.
The corresponding estimators of the quantities $T_c^+$ and  $T_c^-$ are then defined as in Section \ref{sec2}, where
the local linear estimator $\hat \mu _{b_n} $ is  replaced by its
bias corrected version $\tilde \mu _{b_n} $. For example,
the analogue of  the statistic in  \eqref{stat} is given by
\begin{align} \label{statjack}
\tilde{T}_{N,c}^+=\frac{1}{N}\sum_{i=1}^N\int_c^{\infty}\frac{1}{h_d}K_d\Big(\frac{\tilde{\mu}_{b_n}(i/N)-\tilde{\mu}_{b_n}(0)-u}{h_d}\Big)du.
\end{align}
The  investigation of the asymptotic properties of these estimators in the general case  requires some preparations, which are discussed next.

We call a point  $t\in [0,1]$  a {\it regular} point  of the mean function $\mu$,  if the derivative $\mu^{(1)}$ does not vanish at $t$. A point $t\in {\cal C} $ is called a
{\it critical point of $\mu$ of order  $k \geq 1 $}
if the first $k$ derivatives of $\mu$ at $t$ vanish while the $(k+1)$st derivative of $\mu$ at $t$ is non zero,
that is $\mu^{(s)}(t)=0$ for $1\leq s\leq k$ and  $\mu^{(k+1)}(t)\neq0$. Regular points are critical points of
order $0$.
Theorem \ref{Thm1} or  \ref{Thm2} are not valid if any of the roots
of the equation $ \mu(t)-\mu(0) =c$ or $ \mu(t)-\mu(0) =-c$ is a  critical point of order larger or equal than $1$.
The following result provides the asymptotic distribution in this case and also solves the bias problem mentioned  in Section \ref{sec3}.
For its statement we make the following additional assumptions. 

\medskip
\begin{assumption}\label{A4.1} {\rm 
The mean function $\mu $ is three times continuously differentiable. Let
   $t_1^+ , \ldots , t^+_{k^+} $  and  $t_1^- , \ldots , t^-_{k^-} $ denote  the roots of
    the equations $\mu(t)-\mu(0)=c$ and
    $\mu(t)-\mu(0)=-c$, respectively. For each $t_s^-$ ($s=1,\ldots , k^-$) and each 
     $t_s^+$ ($s=1,\ldots , k^+$) there exists a neighbourhood of $t_s^-$  and $t_s^+$ such that 
  $\mu$ is $(v^-_s+1)$ and $(v^+_s+1)$ times  differentiable in these neighbourhoods  with corresponding critical order $v_s^-$ and $v_s^+ $, respectively
 $(1\leq s\leq k^-, ~1\leq s\leq k^+)$. We also assume that the $(v^-_s+1)$st and $(v^+_s+1)$st derivatives of the mean function 
 are Lipschitz continuous on these neighbourhoods. 
 }
\end{assumption}

\begin{assumption} \label{A4}
{\rm There exist $q$ points $0=s_0 < s_1 <
  \ldots < s_q<s_{q+1}=1$ such that the mean function $\mu$ is strictly monotone on each interval  $(s_i,s_{i+1}]$ ($0\leq i\leq q$).}
\end{assumption}

\medskip
\noindent 
It is shown in Lemma \ref{sizeset} of the online supplement that under the assumptions made so far
the set $\{t:|\mu(t)-c|\leq h_n,t\in [0,1]\}$ can
be decomposed as a union of  disjoint ``small'' intervals around the critical  roots $t_i^{+}  $ and  $t_i^{-} $,  whose  Lebesgue measure  is of order
$h_n^{1 / {( v_i^+}+1) }$ and $h_n^{1/ ({v_i^-}+1)}$, respectively,
and therefore depends on the order of the corresponding  root. In the appendix we prove the following result, which clarifies the distributional  properties of
the estimator $\tilde  T_{N,c}^{+}$ defined in \eqref{statjack}
if the sample size converges to infinity.

 \begin{theorem}\label{criticalversion}
Suppose that $k^+\geq 1$, and that  Assumptions \ref{A1}, \ref{A2}, \ref{A4.1} and Assumption \ref{A4}
are  satisfied. Define $v^+=\max_{1\leq l\leq k^+}v_l^+$ as the maximum critical order of the roots of the equation $\mu(t)-\mu(0)= c$ and introduce the notation
 \begin{align}
\chi^+_n& =\Big(b_n^6+\frac{1}{nb_n}\Big)h^{-2}_dh_d^{\frac{1}{v^++1}}
 ~,~~R^+_{1,n}=h_d^{-\frac{v^+}{v^++1}}\Big(b_n^3+\frac{1}{nb_n}\Big)
 ~, \\
 R^{+}_{2,n}& =\frac{n^{1/4}\log^2n}{nb_n}h_d^{-\frac{v^+}{v^++1}}~,~~ R^+_{3,n}=\Big(\frac{1}{Nb_n}+\frac{1}{Nh_d}\Big)\Big(b_n\wedge h_d^{\frac{1}{v^++1}}\Big).
 \end{align}
Assume further that  the bandwidth conditions $h_d\rightarrow 0$, $nb_nh_d\rightarrow \infty$, $b_n\rightarrow 0$, $nb^2_n\rightarrow \infty$, $Nb_n\rightarrow \infty$, $Nh_d\rightarrow \infty$ and $\pi_n=o(h_d)$ hold, where
\begin{align}
\label{b1}&\pi_n:=(b_n^3+(nb_n)^{-1/2}\log n)\log n,
     \end{align}
then we have the following results.
\\
\smallskip
(a) If $b_n^{v^++1}/h_d\rightarrow \infty$,  $\sqrt{nb_n}h_d^{\frac{v^+}{v^++1}}(\chi^+_{n}+R^+_{1,n}+R^+_{2,n}+R^+_{3,n})=o(1)$, 
$\sqrt{nb_n}h_d^{\frac{v^+}{v^++1}}/N=o(1)$, then
\begin{align}
\sqrt{nb_n}h_d^{\frac{v^+}{v^++1}}\left(\tilde T_{N,c}^+-T_{c}^+\right)\convd {\cal N} (0, \sigma_1^{2,^+}+\sigma_2^{2,^+} ) ,
\end{align}
where
\begin{align}\ \ \sigma_1^{2,^+} =\Big(\int K_d(z^{v^++1})dz\Big)^2((v^++1)!)^{\frac{2}{v^++1}}\sum_{\{t_l^+\colon v_l^+=v^+\}} \frac{\sigma^2(t_l^+)}{|\mu^{(v^++1)}(t_l^+)|^{\frac{2}{v^++1}}}
\int (K^*(x))^2dx, \label{DefinitionU4a}\\
\sigma_2^{2,^+}=\sigma^2(0)((v^++1)!)^{\frac{2}{v^++1}}\int (\bar K^*(t))^2dt
\Big (\sum_{\{t_l^+\colon v_l^+=v^+\}} |\mu^{(v^++1)}(t_l^+)|^{ \frac{-1}{v^++1}}
\int K_d(z^{v^++1})dz\Big )^2.\label{DefinitionU4}
\end{align}
(b) If $b_n/h_d^{\frac{1}{v^++1}}=r\in [0,\infty)$, $\sqrt{nh_d}h_d^{\frac{v^+}{2(v^++1)}}(\chi^+_{n}+R^+_{1,n}+R^+_{2,n}+R^+_{3,n})=o(1)$, then
\begin{align}
\sqrt{nh_d}h_d^{\frac{v^+}{2(v^++1)}}\Big(\tilde T^+_{N,c}-T_{N,c}^+\Big)\convd  {\cal N} (0, \rho_1^{2,+}+\rho_2^{2,+} ) ,
\end{align}
 where
\begin{align}
\label{rootsimu}
\rho_1^{2,+}
&=|(v^++1)!|^{\frac{1}{v^++1}}\sum_{\{t_l^+\colon v_l^+=v^+\}}\frac{\sigma^2(t^+_l)}{|\mu^{(v^++1)}(t_l^+)|^{\frac{2}{v^++1}}}
\int\int\int K^*(u)K^*(v)K_d(z_1^{v^++1})\notag
\\&
\times K_d\Big (\Big (z_1+r\Big |\frac{(v^++1)!}{\mu^{(v^++1)}(t^+_l)}\Big |^{\frac{-1}{v^++1}}(v-u)\Big )
^{v^++1}\Big )dudvdz_1,
\end{align}
and $ \rho_2^{2,+} =r^{-1}\sigma_2^{2,+}$,
where $\sigma_2^{2,+}$ is defined in \eqref{DefinitionU4}
\end{theorem}

\medskip
\noindent
 In general the rate of convergence of the estimator
$\tilde T_{N,c}^+$ is determined by the maximal order of the  critical  points, and only critical
points of maximal order appear in the  asymptotic variance. The rate of convergence additionally
depends on the relative order of  the  bandwidths $b_n$ and $h_d$.
 Theorem \ref{criticalversion} also covers  the case $v^+=0$, where all  roots of the equation $\mu (t) - \mu (0) =c$
are regular.
Moreover, the use of the Jackknife corrected estimate $\tilde \mu_{b_n}$ avoids the bias problem observed in Theorem \ref{Thm1}.

 It is also worthwhile to mention that there exists a slight difference in the statement of 
 part  (a) and (b) of  Theorem \ref{criticalversion}.
While  part (a) gives the asymptotic distribution of
$\tilde T^+_{N,c}-T^+_c$ (appropriately standardized),
part  (b) describes the  weak convergence of $\tilde T^+_{N,c}-T^+_{N,c}$. The replacement of $T^+_{N,c}$ by its limit $T^+_c$
is only possible under additional bandwidth conditions. In fact, if
 $b_n/h_d^{\frac{1}{v^++1}}=r\in [0,\infty)$, Theorem \ref{criticalversion} and Proposition \ref{prop1}
 give
\begin{align}
\sqrt{nh_d}h_d^{\frac{v^+}{2(v^++1)}}\Big(\tilde T^+_{N,c}-T^+_c\Big)-R_n\convd {\cal N} (0, \rho_1^{2,+}+\rho_2^{2,+} ),
\end{align}
where $\rho_1^{2,+}$ and $\rho_2^{2,+}$ are defined in Theorem  \ref{criticalversion}, and $R_n$ is a
an additional bias term of order
$$
O(\sqrt{nh_d}h_d^{\frac{v^++2}{2(v^++1)}}),
$$
which does not necessarily vanish asymptotically. For example, in the regular case  $v^+=0$  this bias  is of order $o(1)$
under the additional assumptions $nh_d^3=o(1)$ and $b_n/h_d<\infty$. Note that these bandwidth conditions do not allow for  the MSE-optimal bandwidth  $b_n \sim n^{-1/5}$.  These  considerations
give some arguments for using small bandwidths $h_d$ in the estimator \eqref{statjack} such that condition
(a) of Theorem \ref{criticalversion} holds, that is $h_d = o(b_n^{v^+ +1})$. Moreover, in numerical experiments we observed that smaller bandwidths $h_d$ usually yield a substantially better performance of the estimator
$\tilde T_{N,c}^+$ and in the remaining part of this section we concentrate on this case as this is
most important from a practical point of view.

The next result gives a corresponding statement of the joint asymptotic distribution of $(\tilde T_{N,c}^+,\tilde T_{N,c}^-)$ and as a consequence that  of
$\tilde T_{N,c} = \tilde T_{N,c}^+ + \tilde T_{N,c}^-$,
where the statistic $\tilde T^-_{N,c}$ is defined by
\begin{align} \label{stat-}
\tilde{T}^-_{N,c}=\frac{1}{N}\sum_{i=1}^N\int_{-\infty}^{-c}\frac{1}{h_d}K_d\left(\frac{\tilde{\mu}_{b_n}(i/N)-\tilde{\mu}_{b_n}(0)-u}{h_d}\right)du.
\end{align}

 \begin{theorem}\label{Thm4}
 Assume that the  conditions of Theorem \ref{criticalversion} are satisfied, that $k^-\geq 1$
 and define $v^- = \max_{1 \leq l \leq k^-}  v^-_l$ as the maximum order of the critical roots
 $\{t_l^- \colon 1\leq l\leq k^-\}$. If, additionally, the bandwidth conditions  (a) of Theorem \ref{criticalversion} hold
 and similar bandwidth conditions are satisfied  for the level $-c$,  we have
 \begin{align}
 \sqrt{nb_n} \Big (h_d^{\frac{v^+}{v^+1}}(\tilde T_{N,c}^+-T_c^+),h_d^{\frac{v^-}{v^-+1}}(\tilde T_{N,c}^- -T_c^-) \Big )^T\Rightarrow \mathcal{N}(0,\Sigma),
 \end{align}
 where the matrix $\Sigma = (\Sigma_{ij})_{i,j=1,2}$ has the  entries
 $ \Sigma_{11} =\sigma^{2,+}_1+\sigma^{2,+}_2$, $\Sigma_{22}=\sigma^{2,-}_1+\sigma^{2,-}_2,$
 \begin{align*}
 \Sigma_{12} & =\Sigma_{21}=-\sigma^2(0)((v^++1)!)^{\frac{1}{v^++1}}((v^-+1)!)^{\frac{1}{v^-+1}}\int_0^1 (\bar K^*(t))^2dt\notag
 \\
& \times  \sum_{\{t_l^+\colon v_l^+=v^+\}}{  \int K_d(z^{v^++1})dz \over  |\mu^{(v^++1)}(t^+_l)|^{{1}/({v^++1})}}
 \sum_{\{t_l^-\colon v_l^-=v^-\}}
 {\int K_d(z^{v^-+1})dz \over |\mu^{(v^-+1)}(t_l^-)|^{{1}/({v^-+1})}} ,
 \end{align*}
 and  $\sigma^{2,-}_1$, $\sigma^{2,-}_2$ are defined similarly as 
$\sigma^{2,+}_1$, $\sigma^{2,+}_2$ in
\eqref{DefinitionU4a}, \eqref{DefinitionU4}, respectively.
 \end{theorem}

 \medskip

\noindent The continuous mapping theorem and
Theorem \ref{Thm4} imply the weak convergence of the
estimator $\tilde T_{N,c}$ of  $T_c$, that  is
$
	\sqrt{nb_n}h_d^{\frac{v}{v+1}}(\tilde T_{N,c}-T_c)\rightarrow N(0,\sigma^2),
$
    where  $v=\max\{v^+,v^-\}$ and the
    asymptotic variance is given by 
    $ \sigma^2=\Sigma_{11}\mathbf 1(v^+\geq v^-)+\Sigma_{22}\mathbf 1(v^+\leq v^-)+2\Sigma_{12}\mathbf 1 (v^+=v^-). $
\medskip

\subsection{Bootstrap}  
\label{sec42}

Although Theorem \ref{criticalversion} is interesting from a theoretical point of view and avoids the bias problem described in Section \ref{sec3},
it can not be easily used to construct a test for the hypotheses \eqref{nullnew}.
The asymptotic variance of the statistics $T_{N,c}^+$ and $T_{N,c}^-$
depends on the long-run variance $\sigma^2(\cdot)$ and the set ${\cal C}$ of critical points, which are difficult to estimate.
Moreover, the  order of the critical  roots is usually unknown and not estimable. Therefore it is not clear
which derivatives have to be estimated (the estimation of higher order
derivatives of the mean function is a hard problem anyway). 
As an alternative we propose a bootstrap test which
does not require the estimation of the derivatives of the mean trend at the critical roots.

{
The bootstrap procedure is motivated by an
essential step in the proof of Theorem \ref{criticalversion}, which gives
a stochastic approximation for the difference 
\begin{align*}
\tilde  T_{N,c}^+ -T_c^+&= I'
+ o_p\Big (\big(\sqrt{nb_n}h_d^{\frac{v^+}{v^++1}}\big)^{-1}\Big ),
\notag
\end{align*}
where the statistic $I'$ is defined as 
\begin{align} \label{iprime}
 \ \ \ \ \ \frac{-1}{nNb_nh_d}\sum_{j=1}^n\sum_{i=1}^{ N}K_d\Big (\frac{\mu(i/N)-\mu(0)-c}{h_d}\Big ) \sigma
\big (\frac{j}{n}\big )\Big(K^*\Big (\frac{i/N-j/n}{b_n}\Big)-\bar K^*\Big(\frac{j}{nb_n}\Big)\Big)V_j,
\end{align}
and $(V_j)_{j \in \N}$ is a sequence of independent standard normally distributed random variables.
Based on this approximation  we propose the following bootstrap to calculate critical values.

\medskip
 \begin{algorithm}\label{Algorithm_6.1}{\rm  ~~\\
(1) { Choose bandwidths $b_n$, $h_d$ and an  estimator of the long-run variance,  say $\hat \sigma^2(\cdot)$, which is uniformly consistent on the set
    $
    \cup_{k=1 }^{v ^+} ~ {\cal U}_\varepsilon (t_k^+)
    $
    for some $\varepsilon >0$, where $ {\cal U}_\varepsilon  (t) $ denotes a $\varepsilon$-neighbourhood of the point $t$.} \\
    \smallskip
(2) Calculate   the bias corrected local linear estimate  $\tilde \mu_{b_n}(t)$  and the statistic $\tilde {T}^+_{N,c}$
    defined in \eqref{Jack} and \eqref{statjack}, respectively. \\
    \smallskip
(3) Calculate
      \begin{align} \label{vbar}
    \bar V=\sum_{j=1}^n\hat \sigma^2\big ( \frac{j}{n} \big )\Big [
 		\sum_{i=1}^NK_d\Big (\frac{\tilde \mu_{b_n}(i/N)-\tilde \mu_{b_n}(0)-c}{h_d}\Big )\Big \{ K^*\Big (\frac{i/N-j/n}{b_n}\Big)-\bar K^*\Big(\frac{j}{nb_n}\Big)\Big\}
        \Big ]^2.
\end{align}
(4) Let $q_{1-\alpha}^+$ denote the  the $1-\alpha$ quantile of
    a centered normal distribution with variance $\bar V$, then the
    null hypothesis in \eqref{nullnew+} is rejected, whenever
    \begin{align}\label{boottest}
 nNb_nh_d\big (\tilde T_{N,c}^+-\Delta \big )>q_{1-\alpha}^+.
\end{align}	
 	}
 \end{algorithm}

 \begin{theorem}\label{algorithm_5.1}
Assume that the conditions of Theorem \ref{criticalversion}  (a) are satisfied, then the test \eqref{boottest} defines
a consistent and asymptotic level $\alpha$ test for the hypotheses \eqref{nullnew+}.
\end{theorem}

\begin{remark} \label{rempower}
{\rm ~~ \\
(a)
It follows from  the proof of Theorem \ref{algorithm_5.1} in the appendix
that
\begin{align}
  \mathbb{P} \big ( \mbox{ test } \eqref{boottest} \mbox{  rejects } \big ) \longrightarrow
 \left\{ \begin{array}{cc}
 1 &  \mbox{ if } ~T_c^+ >  \Delta\\
  \alpha &  \mbox{ if } T_c^+ = \Delta\\
   0 &  \mbox{ if } ~T_c^+ <  \Delta\\
 \end{array}
 \right. ~.
\end{align}
 
Moreover, these arguments also show that the power of the test
\eqref{boottest}
depends on the ``signal to noise ratio''  $(\Delta -T_c^+)/\sqrt{\sigma_1^{2,+} +  \sigma_2^{2,+}}$
and that it 
is able to detect local alternatives converging to the null at a rate $O ((nb_n)^{-1/2}h_d^{-{v^+}/(v^++1)})$.  When the level $c$ decreases, the value of $T_c^+$ increases and
 the rejection probabilities also increase. On the other hand, for any given level $c$, the rejection probability will increase when the threshold $\Delta$ decreases (see equation  \eqref{new.5.6} in the appendix). 
 \\

(b) As pointed out by one referee, it is also of interest to discuss some uniformity properties
in this context. For this purpose we consider the  situation in Theorem \ref{algorithm_5.1}, assume
that $f$ is a potential mean function in \eqref{protype} and denote by $v^+_f$ and $q_f$ 
the corresponding quantities in Assumption \ref{A4.1} and \ref{A4} for $\mu =f$.
For given numbers $\tilde q, \tilde v < \infty$ let $\mathcal F$ denote the class  
of all  $3\vee (\tilde v +1)+1$ times differentiable functions $f$  on the interval $[0,1]$
satisfying $\sup_{f \in\mathcal F} v^+_f \leq \tilde v $ and 
$\sup_{f \in\mathcal F} q_f \leq \tilde q $.
Consider a  sequence $(\Delta_n)_{n\in \N} $ satisfying 
$$
\sqrt{nb_n}h_d^{\frac{\tilde v}{ \tilde v +1}}(\Delta- \Delta_n)\rightarrow -\infty
$$
and define for a given level $c>0$, constants $M$, $L$, $\eta, \iota >  0$ the set  $\mathcal F_c(M,\eta, \iota, \tilde q,\tilde v, L,  \Delta_n)$ as the  class of all functions  $ f \in \mathcal F$ with the properties 
\begin{description}
\item (i) The cardinality of the set 
$\mathcal{E}_c^+ (f) = \{ t \in [0,1] : f(t)-f(0)=c \} $ is  at most $M$.
\item (ii) 
$ \min \{ |t_1 - t_2 | : t_1,t_2 \in \mathcal{E}_c^+ (f) ; t_1 \not = t_2 \} \geq \eta$; 
$ \min \{ t_1  : t_1 \in \mathcal{E}_c^+ (f) \} \geq \eta $; \\ $ \max \{ t_1  : t_1\in \mathcal{E}_c^+ (f)  \} \leq  1-\eta  $.  
\item (iii) 
$\sup_{t\in [0,1]  } (f(t)-f(0) ) \geq c+\iota $.
\item (iv) 
$\sup_{t\in [0,1]}\max_{1\leq s\leq 3\vee (\tilde v+1)+1}|f^{(s)}(t)|\leq L$.
\item (v)  $T^+_{f,c}:=\int \mathbf 1(f(t)-f(0)> c)dt\geq  \Delta_n.$
\end{description}
If $\mathbb{P}_f$ denotes the  distribution of the  process $(X_{i,n})_{i=1,\ldots , n}$ in model \eqref{protype} with $\mu=f$,
then  it follows by a careful inspection of the  proof of Theorem \ref{algorithm_5.1}  
that 
\begin{align}
\lim_{n\rightarrow \infty} ~
\inf_{f\in \mathcal F_c(M,\eta, \iota, \tilde q,\tilde v , L,  \Delta_n) }\mathbb{P}_f \big ( \mbox{ test } \eqref{boottest} \mbox{  rejects  } \big ) =1 .
\notag
\end{align}\\
(c)
The bootstrap procedure can easily be modified to test the hypothesis \eqref{nullnew} referring to the quantity $T_c$. 
In step (2), we additionally calculate the
statistic  $\tilde {T}^-_{N,c}$ defined in \eqref{stat-}, $\tilde T_{N,c}  = \tilde T^+_{N,c} + \tilde T^-_{N,c}$
and the quantity
$$ V^*=\sum_{j=1}^n\hat \sigma^2(j/{n})\Big (
 	\sum_{i=1}^NK^\dag_d\Big (\frac{\tilde \mu_{b_n}(i/N)-\tilde \mu_{b_n}(0)-c}{h_d}\Big )\Big (K^*\Big (\frac{i/N-j/n}{b_n}\Big)-\bar K^*\Big(\frac{j}{nb_n}\Big)\Big)\Big )^2,$$
 	where $$
 	K^\dag_d\Big (\frac{\tilde \mu_{b_n}(i/N)-\tilde \mu_{b_n}(0)-c}{h_d}\Big )=K_d\Big (\frac{\tilde \mu_{b_n}(i/N)-\tilde \mu_{b_n}(0)-c}{h_d}\Big )-K_d\Big (\frac{\tilde \mu_{b_n}(i/N)-\tilde \mu_{b_n}(0)+c}{h_d}\Big ).$$
 	Finally, the null hypothesis \eqref{nullnew} is rejected if
 	$ nNb_nh_d\big (\tilde T_{N,c}-\Delta \big )>q_{1-\alpha},$
    where
 	$q_{1-\alpha}$ denotes the $(1-\alpha)$th quantile of
    a centered normal distribution  with variance  $V^*$。
 }
    \end{remark}

\medskip
\noindent
For the estimation of the  the long-variance 
we define $S_{k,r}=\sum_{i=k}^rX_i$ and for $m\geq 2$ 
$$
\Delta_j=\frac{S_{j-m+1,j}-S_{j+1,j+m}}{m},
$$
and for $t\in[m/n,1-m/n]$
\begin{align}\label{2018-5.4}
\hat \sigma^2(t)=\sum_{j=1}^n\frac{m\Delta_j^2}{2}\omega(t,j),
\end{align} where for some bandwidth $\tau_n\in(0,1)$, $$
\omega(t,i)=K\Big(\frac{i/n-t}{\tau_n}\Big)/\sum_{ {i}=1}^nK\Big(\frac{i/n-t}{\tau_n}\Big).
$$
For $t\in[0,m/n)$ and   $t\in(1-m/n,1]$ we define
$\hat \sigma^2(t)=\hat \sigma^2(m/n)$  and 
$\hat \sigma^2(t)=\hat \sigma^2(1-m/n)$, respectively.
Note that  the estimator \eqref{2018-5.4}  does not involve 
estimated residuals. The following result shows that  $\hat \sigma^2 $  is consistent and 
can be used in Algorithm \ref{algorithm_5.1}.

\begin{theorem} \label{Thm6_new}
Let Assumption \ref{A1} - \ref{A2} be satisfied and assume 
$\tau_n\rightarrow 0, n\tau_n\rightarrow \infty$,  $m\rightarrow \infty$ and $\frac{m}{n\tau_n}\rightarrow 0$.
 If, additionally, the function $\sigma^2$ is twice continuously differentiable,  then the estimate defined in \eqref{2018-5.4}
 satisfies
	\begin{align}
	\sup_{t\in [\gamma_n,1-\gamma_n]}|\hat \sigma^2(t)-\sigma^2(t)|=O_p\Big(\sqrt{\frac{m}{n\tau_n^2}}+\frac{1}{m}+\tau_n^2+m^{5/2}/n\Big),\label{supt}
    \end{align}
    where $\gamma_n=\tau_n+m/n$. Moreover,  we have
	\begin{align}
	\hat \sigma^2(t)-\sigma^2(t)=O_p\Big(\sqrt{\frac{m}{n\tau_n}}+\frac{1}{m}+\tau^2_n+m^{5/2}/n\Big)\label{fix_t}
	\end{align}
 for any   fixed $t\in(0,1)$,   and for $s=\{0,1\}$
 \begin{align}
	\hat \sigma^2(s)-\sigma^2(s)=O_p\Big(\sqrt{\frac{m}{n\tau_n}}+\frac{1}{m}+\tau_n+m^{5/2}/n\Big).
	\end{align}
	\end{theorem}
\medskip

\noindent 
Note that error term  $\sqrt{\frac{m}{n\tau_n}}+\frac{1}{m}+\tau_n^2$ in \eqref{fix_t} is minimized at the rate of $O(n^{-2/7})$ by $m\asymp n^{2/7}$ and $\tau_n\asymp n^{-1/7}$, where we write $r_n\asymp s_n$ if $r_n=O(s_n)$ and $s_n=O(r_n)$.  For this choice the estimator 
\eqref{2018-5.4} achieves { a better rate than  the long-run variance estimator proposed in 
 \cite{zhou2010simultaneous} (see Theorem 5 in this reference).}

 \section{Simulation study} \label{sec6}
\def\theequation{5.\arabic{equation}}
\setcounter{equation}{0}

In this section we investigate the finite sample properties
of the bootstrap tests proposed in the previous sections. For the sake of brevity we restrict ourselves to the test
\eqref{boottest} for the hypotheses \eqref{nullnew+}. Similar results can be obtained for the corresponding tests for the
hypotheses \eqref{nullnew} and \eqref{nullnew-}.
{The code used to obtain the presented results is available from the second author on request.}

{\color{black}Throughout this section all kernels are chosen as Epanechnikov kernel.} The selection of the  bandwidth  $b_n$ in the local linear estimator is of particular importance in our approach, and for this purpose we use the generalized cross validation  (GCV) method. 
 To be precise, let $\tilde e_{i,b}=X_{i,n}-\tilde \mu_b(i/n)$ be the  residual obtained from a
 bias corrected local linear fit with  bandwidth $b$ and define  $\tilde{\bf e}_b=(\tilde e_{1,b},\ldots ,\tilde e_{n,b})^T$.  Throughout this section we use the bandwidth
$$
\hat b_n=\argmin_b GCV(b):=\argmin_b\frac{n^{-1}\hat{\bf e}^T_b\hat \Gamma_n^{-1}\hat{\bf e}_b}{(1-K^*(0)/(nb))^2}~,
$$
where $\hat \Gamma_n$ is an estimator of the covariance matrix $\Gamma_n:=\{\E(\epsilon_{i,n}\epsilon_{j,n})\}_{1\leq i,j\leq n}$, which is obtained by the banding techniques as described in \cite{wu2009banding}.

It turns out  that Algorithm  \ref{algorithm_5.1}   is not very
sensitive with respect to the choice of the bandwidth $h_d$ as long as it is chosen sufficiently small. Similarly, the number $N$ of knots used in the Riemann approximation \eqref{stat} has a negligible influence on the test, provided  it has been chosen sufficiently large.
As a rule of thumb  satisfying the bandwidth conditions of Theorem \ref{criticalversion}(a), we use $h_d=N^{-1/2}/2$ throughout this section, and investigate the influence of  other choices  below.
The number of knots is always given by $N=n$.
In order to save computational time we use  $m=\lf n^{2/7}\rf$  and $\tau_n= n^{-1/7}$ 
for the estimator $\hat  \sigma^2$ in the simulation study [see the discussion at the end of Section \ref{sec42}]. For the data analysis 
in Section \ref{sec62}  we suggest a data-driven procedure and use a slight modification of the minimal volatility method as proposed by \cite{zhou2010simultaneous}. To be precise - in order  to avoid choosing too large values for  $m$ and $\tau$ -  we penalize the quantity $$ISE_{h,j} =ise[\cup_{r=-2}^2 \hat  \sigma^2_{m_h,\tau_{j+r}}(t)\cup_{r=-2}^2\hat  \sigma^2_{m_{h+r},\tau_{j}}(t)]$$ 
in their selection criteria by the term $2(\tau_j+m_h/n)IS$, where 
$\hat  \sigma^2_{m_h,\tau_j}(\cdot)$ is the estimator \eqref{2018-5.4} of the long-run variance with parameters $m_h$ and $\tau_j$ and 
$IS$ is the average of the quantities $ISE_{ {h},j} $.

\begin{figure}[t]
\centering
\includegraphics[width=9cm,height=7.5cm]{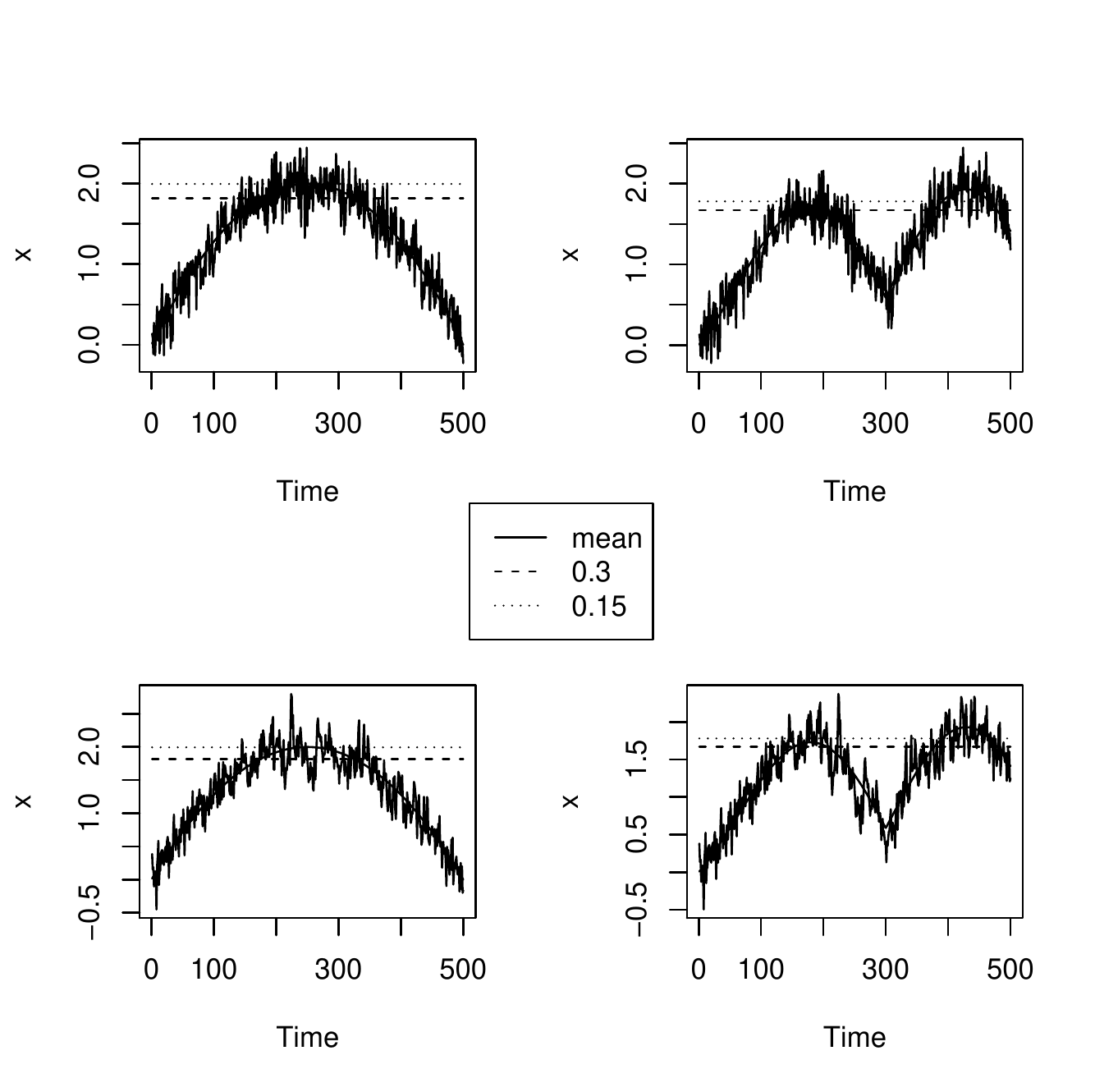}
\vspace{-.4cm}
\caption{\it Simulated sample paths for the four models under consideration. The horizontal lines
display the level $c$ which is given by  $1.82$ and $1.995$
for the mean  function (a) and  by $1.672$ and $1.78$
 for the mean function (b).}
 \label{powerplot1}
\end{figure}


All simulation results presented in  this section are based on $2000$ simulation runs.
We consider the
model \eqref{protype} with errors $\epsilon_{i,n} = G(i/n,\FF_i)/5$, where
\begin{description}
	\item (I)  {:} $G(t,\FF_i)=0.25|\sin(2\pi t)|G(t,\FF_{i-1})+\eta_i$;
	\item  (II)  {:} $G(t,\FF_i)=0.6(1-4(t-0.5)^2)G(t,\FF_{i-1})+\eta_i$ {,}
\end{description}
and the filtration  $ \FF_i=(\eta_{-\infty},\ldots ,\eta_i)$ is  generated by a sequence
 $\{\eta_i,i\in \mathbb Z\}$ of independent standard normally distributed random variables.
For the  mean trend we consider the  following two cases
\begin{description}
\item (a): $\mu(t)=8(-(t-0.5)^2+0.25);$
\item (b): $\mu(t)=\sin( 2|t-0.6|\pi)(1+0.4t)$.
\end{description}
{
Typical sample paths of these processes are depicted  in Figure \ref{powerplot1}.
Note that the mean trend (b) is not differentiable at the point $0.6$. However, using similar but more complicated arguments as given in Section \ref {sec7} and in the supplementary material, it can be shown that the results of this paper also hold if $\mu(\cdot) $ is Lipschitz continuous outside of an open set containing the critical 
roots  $ t^+_1, \ldots  , t^+_{k^+}, t^-_1, \ldots ,  t^-_{k^-}$.}

{We begin illustrating the finite sample properties of the (uncorrected) estimator $\hat{T}_{N,c}^+$  in \eqref{stat}
and its bias correction  $\tilde T_{N,c}^+$  in 
\eqref{statjack} for the quantity $T_c^+$, where $c=1.8$.
The corresponding values of $T_c^+$ are 
$T_{1.8}^+=0.3163$ and  $T_{1.8}^+=0.1406$ in models (a) and (b), respectively.
In  Table \ref{Tablestimate} we display the bias and  standard deviation 
of the 
two estimators. 
We observe a substantial reduction of the bias  by a factor between
$5$ and $75$, while there is a slight increase in standard deviation. Except for one case the bias corrected estimate  $\tilde T_{N,c}^+$ has a smaller mean squared error than the uncorrected estimate. 
\begin{table}[htbp]
	\centering
	\caption{\it Simulated bias and standard deviation of the estimators 
    $\hat{T}_{N,c}^+$   and its bias correction  $\tilde T_{N,c}^+$, where $c=1.8$.
		The sample size is  $n=500$ and the  bandwidth has been 
        chosen by GCV. }
   \begin{tabular}{|c|cc|cc|cc|cc|}
          \hline
    Model & \multicolumn{2}{c|}{(a,I)} & \multicolumn{2}{c|}{(a,II)} & \multicolumn{2}{c|}{(b,I)} & \multicolumn{2}{c|}{(b,II)} \\
    \hline
    Accuracy & {bias} & {sd} &{bias} & {sd} &{bias} & {sd} & {bias} & {sd} \\
    \hline
    $\hat T^+_{N,1.8}$ & -0.105 & 0.063 & -0.122 & 0.077 & -0.077 & 0.055 & -0.054 & 0.060 \\
    $\tilde T^+_{N,1.8}$ & -0.008 & 0.065 & -0.011 & 0.069 & -0.001 & 0.076 & 0.010  & 0.085 \\
    \hline
    \end{tabular}
	\label{Tablestimate}%
\end{table}
}

{
Next we investigate the finite sample properties of the bootstrap test 
\eqref{boottest} for the hypotheses \eqref{nullnew+}, where the threshold is given by $\Delta=0.3$ and $\Delta=0.15$.}
Following the discussion in Remark \ref{rempower}(a) we display in
 Tables \ref{label-1}  the
 simulated type 1 error at the boundary of the null hypothesis  in \eqref{nullnew+}, that is $T_c^+ = \Delta$.
 A good approximation of  the nominal level at this
point is  required as the rejection probabilities for $T_c^+ < \Delta $ or $T_c^+ > \Delta $
are usually smaller or  larger than this value,
respectively.
The values  of $c$ corresponding to  $ T_c^+=0.3$ and $T_c^+=0.15$
are given by  $c=1.82$ and $c=1.955$ for the mean function  (a) and by
$c=1.672$ and $ c=1.78$ for the mean function  (b).
 We observe a rather  precise  approximation of the nominal level, which is improved with increasing sample size.   For  the  sample size $n=200$ the GCV method selects the bandwidths $b_{cv}$ for  $0.25$, $0.26$, $0.23$, $0.19$
for the models  $((I),(a) )$,  $((I),(b) )$, $((II),(a) )$, and  $((II),(b) )$, respectively.  Similarly, for the sample size $n=500$ the
 GCV method  selects the bandwidths  $0.2$, $0.17$, $0.21$, $0.14$ for the models  $((I),(a) )$,  $((I),(b) )$, $((II),(a) )$   and  $((II),(b) )$, respectively. In order to study the robustness of the test with respect
to the choice of $b_n$  we investigate the  bandwidths $b_{cv}^-=b_{cv}-0.05, b_{cv}, b_{cv}^+=b_{cv}+0.05$. For this range of bandwidths the approximation of the nominal level is remarkably  stable.

\begin{table}[htbp]
  \centering
  \caption{\it Simulated level of the test \eqref{boottest}  at the boundary of  the null hypothesis \eqref{nullnew+}.
	The sample size is  $n=200$ (upper part) and $n=500$ (lower part) and various bandwidths are considered. The bandwidth $b_{cv}$ is chosen by GCV, and $b_{cv}^-=b_{cv}-0.05$, $b_{cv}^+=b_{cv}+0.05.$ 
    }
    \begin{tabular}{cclrrrrrrrr}
    \toprule
$n$ & \multicolumn{1}{l}{~}  & model & \multicolumn{2}{c}{(a,I)} & \multicolumn{2}{c}{(b,I)} & \multicolumn{2}{c}{(a,II)} & \multicolumn{2}{c}{(b,II)} \\
\midrule
&    \multicolumn{1}{l}{$\Delta$} & $b_n$ & 5\%   & 10\%  & 5\%   & 10\%  & 5\%   & 10\%  & 5\%   & 10\% \\
          \cline{2-11}
&    \multirow{3}[0]{*}{0.3} & $b_{cv}^-$ & 4     & 8.95  & 5.35  & 10.1  & 4.9   & 8.8   & 5.6   & 9.35 \\
&          & $b_{cv}$ & 3.5   & 8.2   & 4.15  & 8.05  & 4     & 8     & 6     & 10.7 \\
$200$~~ &          & $b_{cv}^+$ & 4.15  & 7.6   & 2.85  & 5.3   & 3.75  & 6.85  & 4.85  & 9.15 \\
          \cline{2-11}
&    \multirow{3}[0]{*}{0.15} & $b_{cv}^-$ & 5.45  & 8.75  & 5.8   & 9.25  & 6.9   & 10    & 6.45  & 11.55 \\
&          & $b_{cv}$ & 6.45  & 10.8  & 5.35  & 8.7   & 6.45  & 10.7  & 7.25  & 11.05 \\
&          & $b_{cv}^+$ & 5.65  & 10.05 & 2.45  & 4.55  & 6.4   & 10.15 & 5.75  & 9.95 \\
          \midrule
              \midrule
&    \multirow{3}[0]{*}{0.3} & $b_{cv}^-$ & 5.2   & 9.45  & 5.85  & 10.1  & 5.85  & 10.05 & 5.55  & 9.9 \\
&          & $b_{cv}$ & 4.6   & 9.55  & 5.45  & 9.85  & 5.65  & 9.25  & 6     & 10.1 \\
$500$~~
&          & $b_{cv}^+$ & 5.15  & 9.1   & 5     & 8.95  & 3.65  & 7.15  & 5.45  & 9.85 \\
   \cline{2-11}
 &   \multirow{3}[0]{*}{0.15} & $b_{cv}^-$ & 7.6   & 12.1  & 6.5   & 9.6   & \multicolumn{1}{l}{7.7} & 11.15 & 7.5   & 11.3 \\
    &      & $b_{cv}$ & 6.55  & 11.25 & 5.1   & 9.15  & 7.75  & 12.2  & 5.15  & 9.25 \\
    &      & $b_{cv}^+$ & 6.85  & 10.6  & 4.4   & 7.5   & 6.6   & 11.05 & 4.6   & 8.3\\
    \bottomrule
    \end{tabular}%
  \label{label-1}%
\end{table}

\begin{table}[htbp]
  \centering

\caption{\it Simulated level of the test \eqref{boottest}  at the boundary of  the null hypothesis \eqref{nullnew+} for different choices of the bandwidth $h_d$.
	The sample size is  $n=500$. The bandwidth $b_{cv}$ is chosen by GCV.}
 \begin{tabular}{cclrrrrrrrr}
    \toprule
$n$ & \multicolumn{1}{l}{~}  & model & \multicolumn{2}{c}{(a,I)} & \multicolumn{2}{c}{(b,I)} & \multicolumn{2}{c}{(a,II)} & \multicolumn{2}{c}{(b,II)} \\
\midrule
&    \multicolumn{1}{l}{$\Delta$} & $h_d$ & 5\%   & 10\%  & 5\%   & 10\%  & 5\%   & 10\%  & 5\%   & 10\% \\
          \cline{2-11}
          &    \multirow{3}[0]{*}{0.3}  & $0.0224$ & 4.6   & 9.55  & 5.45  & 9.85  & 5.65  & 9.25  & 6     & 10.1 \\ 
          & & $0.0112 $ & 5.3   & 9.5  & 6.75  & 11.01  & 4.95  & 8.85 & 4.6  & 8.15\\
        $500$~~
&          & $0.0056$& 4.9  & 9.5   & 6.7     & 11.25  & 5.2  & 9.3  & 5.25  & 9.5 \\
   \cline{2-11}
 &   \multirow{3}[0]{*}{0.15} & 0.0224 & 6.55  & 11.25 & 5.1   & 9.15  & 7.75  & 12.2  & 5.15  & 9.25 \\
& & $0.0112$ & 6.1   & 10.25  & 5.7   & 9.35   & \multicolumn{1}{l}{6.4} & 10.95 & 5.45 & 8.75\\
     &      & $0.0056 $ & 7.45  & 12.15  & 6.25   & 10.25   & 7.55   & 11.95 & 6.9   & 11.8\\
    \bottomrule
    \end{tabular}%
    \label{label-2}%
    
 \end{table}

\begin{figure}[htbp]
	\centering
\includegraphics[width=9cm,height=4.5cm]{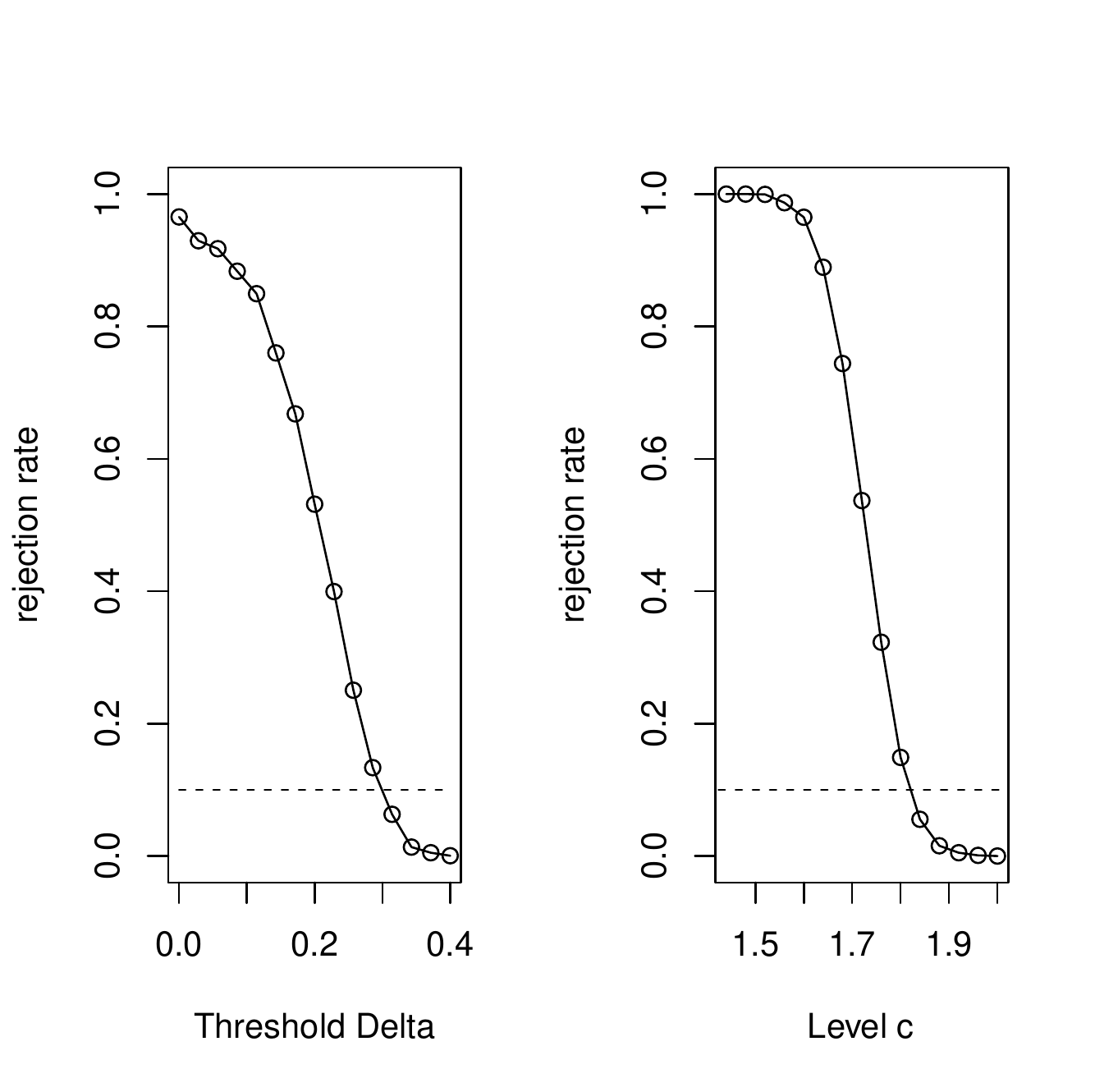}
	\vspace{-.2cm}
    \caption{\it Simulated rejection probabilities of the test \eqref{boottest} 
    in model
	\eqref{protype}
    for varying values of
	$c$ and $\Delta$.
	Left:  $c=1.82$,  $\Delta \in [0,0.4]$ (the case $\Delta=0.3$ corresponds to the boundary of the null hypothesis). Right:
	$\Delta =0.3 $, $c \in [1.44, 2]$ (the case $c=1.82$ corresponds to the boundary of the null hypothesis).The dashed  horizontal line represents the nominal level 10\%.  }\label{powerplot2}
\end{figure}

We also briefly address the problem of the sensitivity of the procedure with respect to the choice of the bandwidth  $h_d$.
For this purpose we consider the same scenarios as in Table \ref{label-1}. For the sake of brevity we restrict ourselves to the case $n=500$ and the data driven bandwidth $b_{cv}$. 
The results are shown in Table \ref{label-2} 
for the bandwidths $h_d=n^{-1/2}/2=0.0224 $,  $h_d= 0.0112 $ and $h_d=0.0056$ and show that the procedure is very stable with respect to the choice $h_d$ as long as $h_d$ is chosen sufficiently small.

In Figure \ref{powerplot2}, we investigate the properties of the test \eqref{boottest} as a function
of the threshold  $\Delta$ and level $c$, where we restrict ourselves to the scenario $((I),(a))$. For the other cases the observations are similar.  The bandwidth  is  $b_n=0.2$.
In the left part of the figure the level $c$ is fixed as $ 1.82$ and $\Delta$ varies  from $0$ to $0.4$ (where
the true threshold is $\Delta=0.3$).  As expected the rejection probabilities decrease  with an increasing threshold $\Delta$.
Similarly, in the right part of  Figure \ref{powerplot2} we display the rejection probabilities for fixed $\Delta=0.3$
when $c$  varies between $1.44$ and  $2$. Again the rejection rates  decrease when $c$ increases.

We finally investigate the power of the test \eqref{boottest} for the hypotheses \eqref{nullnew+}  with
$c=1.82$ and $\Delta=0.3$, where   the bandwidth is chosen as $b_n=0.2$. The  model  is given  by  \eqref{protype} with error  $(I)$  and different mean functions
\begin{align} \label{meanpower}
\mu(t)=a(-(t-0.5)^2+0.25),~~a \in  [7.5,9.5]
\end{align}
are considered (here the case $a=8$ corresponds to the boundary of the hypotheses). 
 The results are presented in Figure \ref{powerplot3}, which demonstrate  that the test \eqref{boottest} has decent power.

\begin{figure}[htbp]
	\centering
	\includegraphics[width=10cm,height=5.4cm]{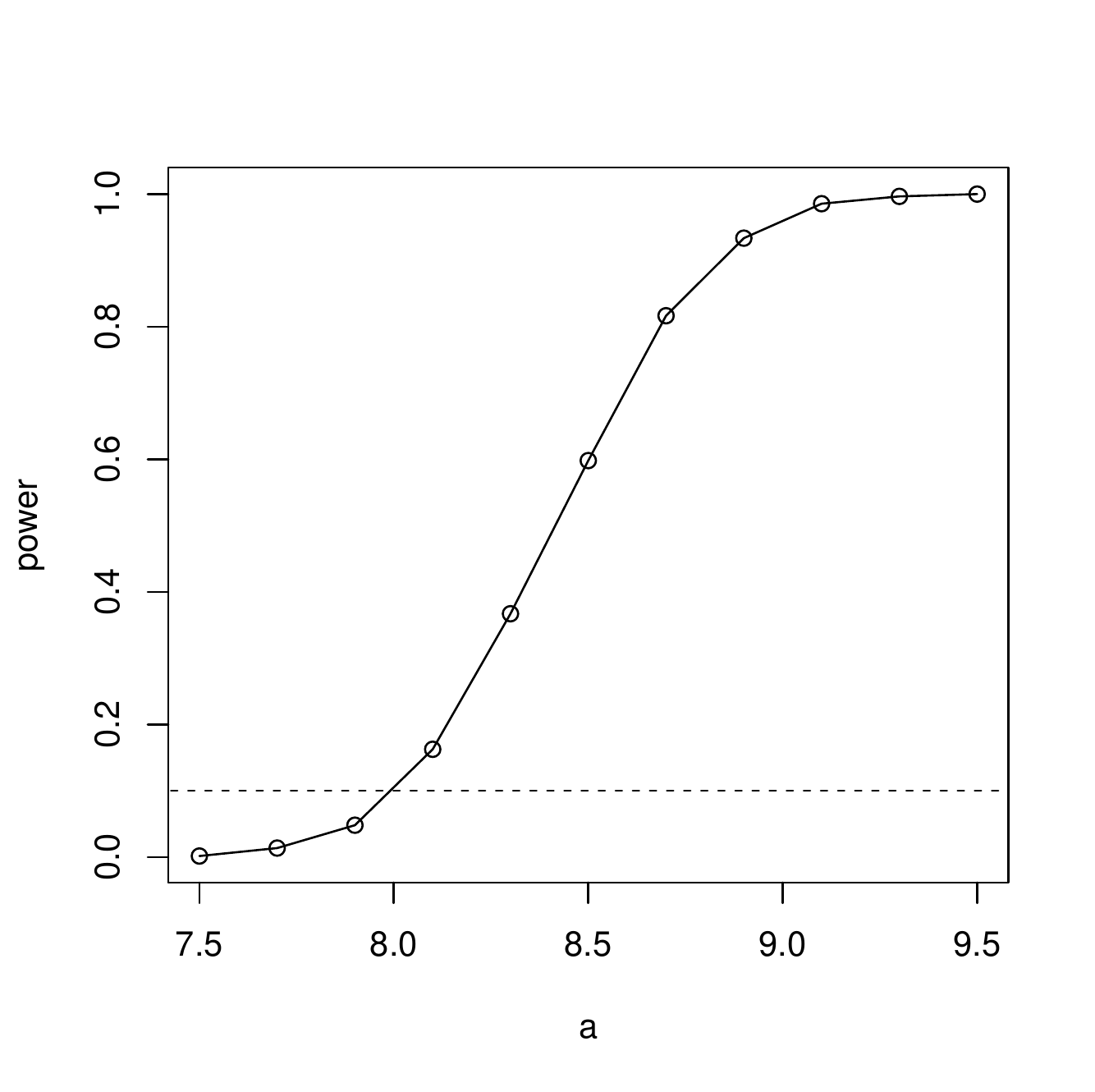}
\vspace{-.1cm}
\caption{\it Simulated power of  the test \eqref{boottest}   in model
	\eqref{protype}
	for the hypothesis \eqref{nullnew+}  with $c=1.82$ and $\Delta=0.3$.
	The mean functions are given by  \eqref{meanpower}  and the case $a= 8$
	corresponds to the boundary of the null hypothesis.
	The dashed  horizontal line represents the nominal level 10\%.	}\label{powerplot3}
\end{figure}

Although hypotheses of the form \eqref{nullnew} have not been investigated in the literature so far it was pointed out by a referee that it might be of interest to see a comparison with tests for similar hypotheses.  The method most similar in spirit to our approach is the test of \cite{DW2016} for  the hypotheses \eqref{nullb}. Note that the procedure 
of these authors assumes a constant mean before and after the (relevant) change point, while  we investigate if  a (inhomogeneous) process deviates  from it's initial mean substantially over a sufficiently long period.
Thus -  strictly speaking - none of the procedures is applicable to the other testing problem. 
\begin{figure}[t]
	\centering
	\includegraphics[width=16cm,height=6.5cm]{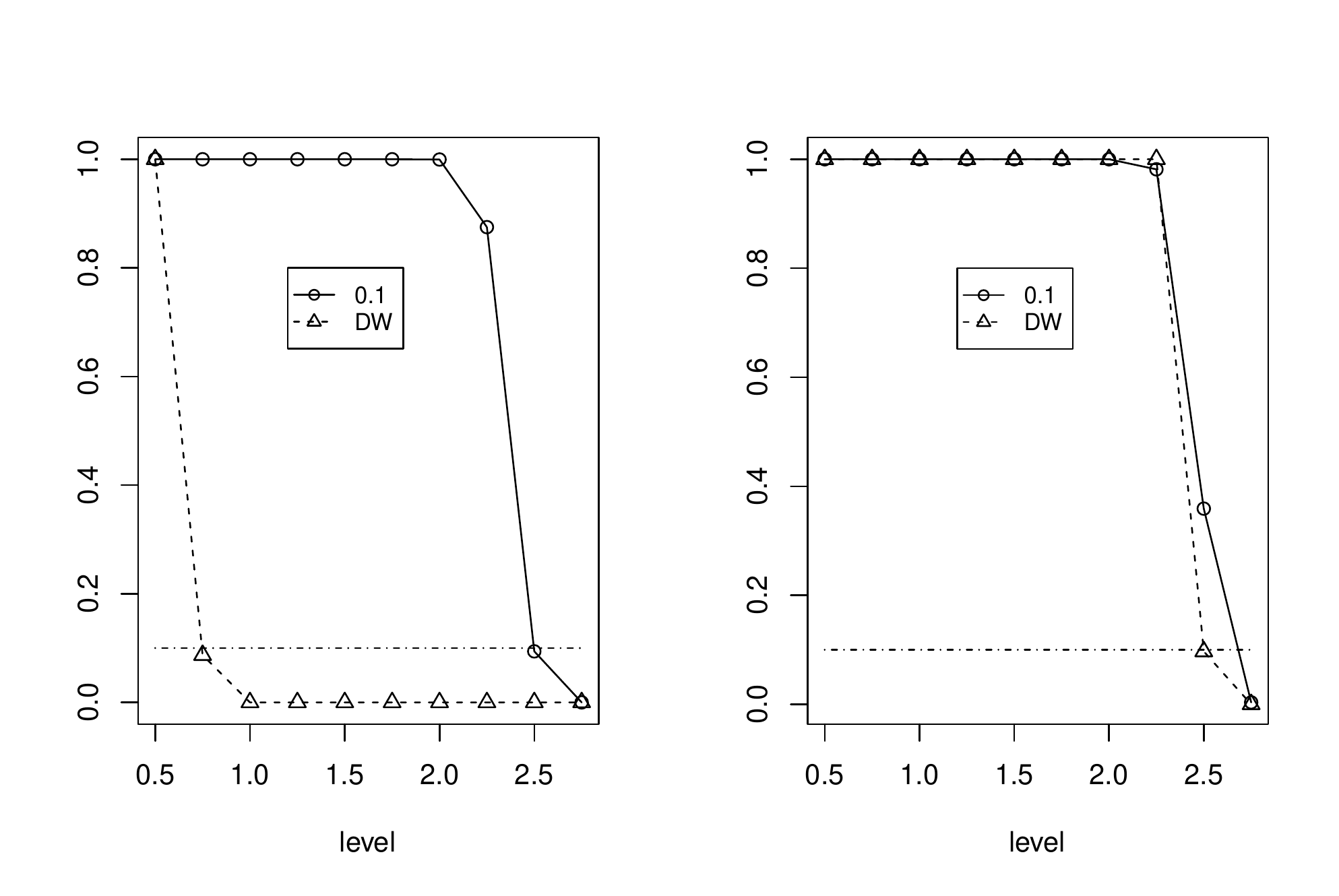}
	\vspace{-.4cm}
	\caption{\it Rejection rates of the test of  \cite{DW2016} (dashed line) and the bootstrap test 
    \eqref{boottest} with $\Delta=0.1$ (solid line) for various values of the level $c$. Left panel: regression function (III); right panel:
    regression function (IV). The nominal level is $10\%$. }
	\label{muans}
\end{figure}
 On the other hand  both tests address the problem of relevant changes under different perspectives
 and it might therefore be of interest to see their performance in the respective alternative testing problems. For this purpose 
 we consider  model \eqref{protype} with the mean functions
 \begin{description}
		\item (III)  $\mu(t)=2.5 \sin(\pi t)$,
        \item (IV) $\mu(t)=0$ for $t\in[0,1/3)$ and $\mu(t)=2.5$ for $t\in[2/3,1]$,
		\end{description}
and an independent error process $ \epsilon_{i,n} \sim N(0,1)/4$. Note that model (III) corresponds to the situation considered in this paper (i.e. a continuously  varying mean function), while model (IV) reflects the situation investigated in \cite{DW2016}. 
In Figure  \ref{muans} we display the rejection probabilities of both tests if the level $c$ varies  
 from $0.5$ to $2.75$ (thus the curves are decreasing with increasing $c$).  The  significance level is given by $10\%$, which means  the 
 value of $c$ where the curve is $10\%$ should be close to 2.5.
 For the hypotheses \eqref{nullnew} we  fixed  $\Delta$ as $0.1$,
 because for a comparison with the test of  \cite{DW2016} it is irrelevant how long the threshold is exceeded and the power of the test \eqref{boottest} decreases
 for increasing values of $\Delta$ (see Figure \ref{powerplot2}).
\\
We observe in the left panel of Figure \ref{muans}  that the test of  \cite{DW2016} performs poorly in model (III),  
where the mean is not constant and the conditions for its applications are not satisfied. 
On the other hand, the bootstrap test \eqref{boottest} shows a reasonable 
performance in model (IV) although the assumptions for its application are not satisfied.
In particular this test   shows a similar performance as the test of  \cite{DW2016} for small values of $\Delta$, which is particularly  designed for
the hypotheses \eqref{nullb} (see the right panel of Figure \ref{muans}). 

\newpage
\section{Data examples}
\label{sec62} 
\def\theequation{6.\arabic{equation}}
\setcounter{equation}{0}

\subsection{Global temperature data} 
 Global temperature data has been extensively studied in the statistical literature under the assumption of  stationarity   [see for example \cite{bloomfield1992climate}, \cite{vogelsang1998trend} and  \cite{wu2007inference}
 among others].  We consider here a series from
http://cdiac.esd.ornl.gov/ftp/trends/temp/jonescru/ with  global monthly temperature anomalies from January  $1850$ to April $2015$, relative to the $1961-1990$ mean. The data and  a local linear estimate of the  mean function  are depicted in left panel of Figure \ref{globaltemperature}.  The figure indicates a non-constant higher order structure of the series and analyzing this series under the assumption of stationarity might be questionable. In fact, the 
test of \cite{dette2015change1} for a constant lag-$1$ correlation 
yields a $p$-value of $1.6\%$ supporting   a non-stationary model for data analysis.

\begin{figure}[htbp]
	\centering
	\includegraphics[width=6cm,height=5cm]{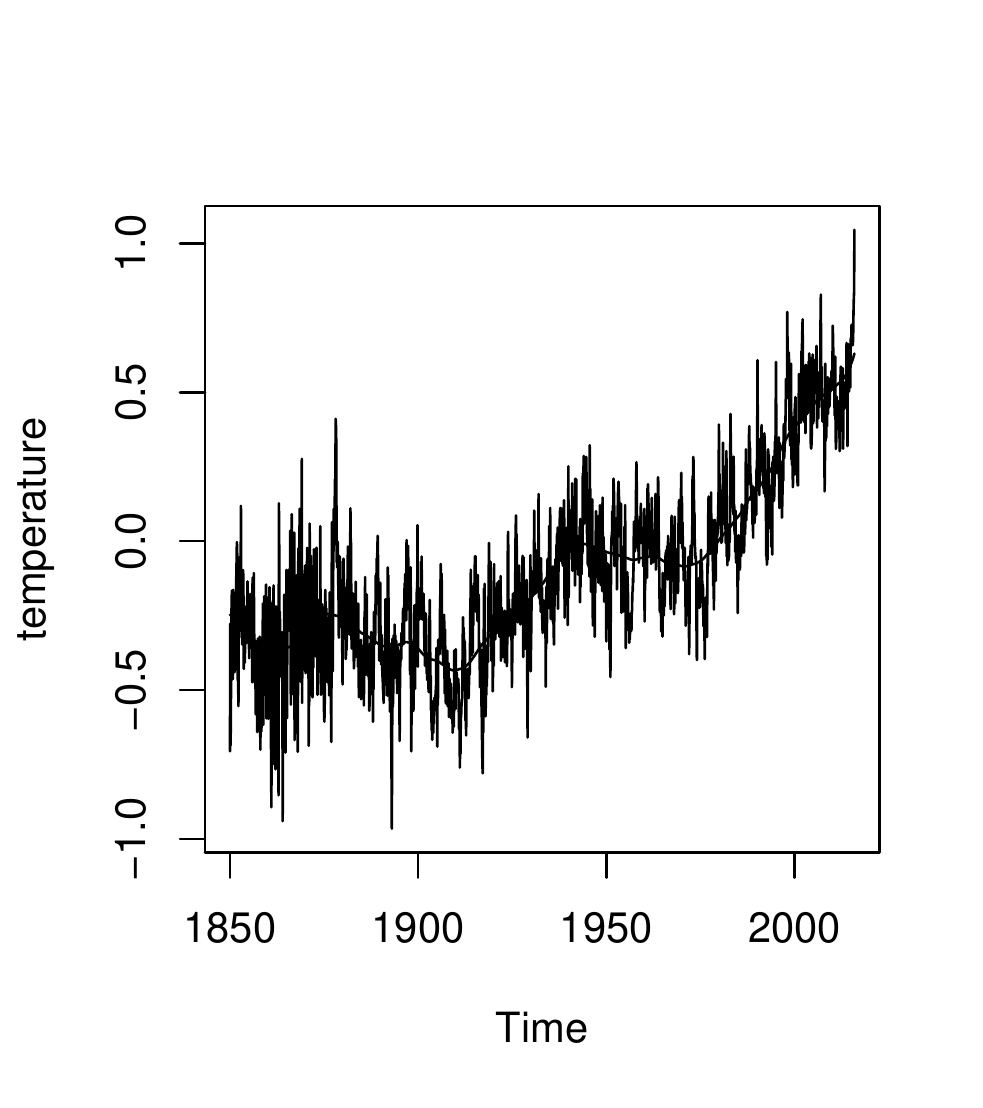}
	\includegraphics[width=6cm,height=5cm]{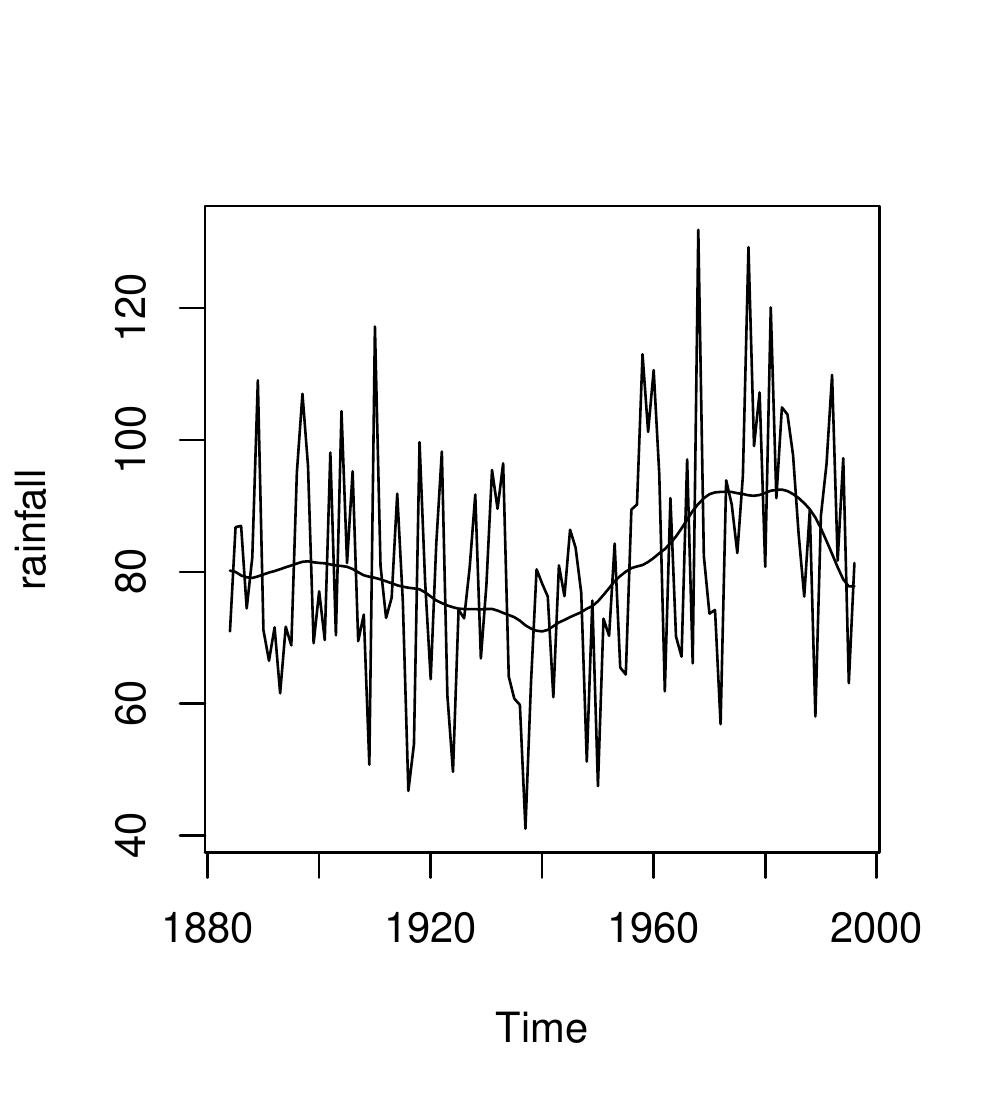}
    \vspace{-.75cm}
	\caption{\it Left panel: deseasonalized global temperature 1850--2015 and its fitted mean-trend.  Right panel: Yearly Rainfall of  Tucum\'{a}n Province, Argentina, 1884--1996. }\label{globaltemperature}
\end{figure}
We are interested in the question if the deseasonalized monthly temperature exceeds the temperature in January $1850$ 
by more than  $c=0.15$ degrees Celsius in more than  $100\Delta \%$ of the considered period. For this purpose  we run the 
test \eqref{boottest} for the hypothesis \eqref{nullnew+}, where the bandwidth (chosen by GCV) is $b_n=0.105$ and $h_d=0.011$ (we note again that the procedure is rather stable with respect
to the choice of $h_d$). For the estimate \eqref{2018-5.4} of the
long-run variance  $\sigma^2$, we use the procedure described at the beginning of this section, which yields $m=30$ and $\tau=0.202$. 
For a threshold $\Delta=43.4\%$ we obtain a  $p$-value of $4.82\%$.

Next we  investigate the same question for  the sub-series from 
January $1850$ to December $1974$.  The GCV method yields  the   bandwidth $b_n=0.135$ 
and we chose  $h_d=0.013$ and  $m=36$, $\tau=0.234$ for the estimate of the time-varying long-run  variance (see the discussion at the beginning of this section).
We find that for $\Delta=26\%$ and $c=0.15$ the $p$-value is $6.6\%$. Comparing the results for the series and sub-series
shows that relevant deviations of more than $c=0.15$ degrees Celsius arise more frequently between $1975$ and $2015$.
The conclusions of this short data analysis are similar to those  of many authors, but by our method we are able to quantitatively describe relevant deviations. For example, if we reject the hypothesis that in less than  $26\%$ of the time between January  $1850$ and April $2015$ the mean function exceeds its value from January 1850 by more than $c=0.15$ degrees Celsius, the type I error of  this conclusion is less or equal than $5\%$.

\subsection{Rainfall data} 
In this example we analyze  the yearly
rainfall data (in millimeters) from $1884$ to $1996$ in the Tucum\'{a}n Province, Argentina, which is a predominantly agriculture region. Therefore its economy well-being depends sensitively on timely rainfall.
The series with a local linear estimate of the mean trend are depicted in right panel of Figure \ref{globaltemperature} (note that the range of estimated mean function is  $[71.0$mm, $92.5$mm]) and it has been studied 
by several authors in the context of change point analysis
with different conclusions.   For example,  the null hypothesis of no change point is rejected by the conventional CUSUM test, isotonic regression approach of \cite{wu2001isotonic} with $p$-value smaller than $0.1\%$, and the 
robust bootstrap test of \cite{zhou2013}
with a  $p$-value smaller than  $2\%$. On the other hand  a self-normalization method considered in \cite{shao2010testing} reports a $p$-value about $10\% $.

Meanwhile, there is some belief that there exists a change point 
because of  the construction of a dam near the region during $1952-1962$. As a result, a more practical question is whether the construction of the dam has a relevant  influence on the economic well-being of the region via affecting the annual rainfall. To investigate this question, we are testing the  hypotheses \eqref{nullnew+} with a threshold $\Delta=0.05$ (here we calculated
$b_n=0.235$, $m=11$, $\tau=0.24$  and $h_d=0.047$ as described at the beginning of this section). For the level  $c=7$ the $p$-value  is $6.03\%$.
In other words the hypothesis that in less than 
$5\%$ of the $113$ years the mean annual rainfall is at least $7$mm higher than the rainfall in the year $1880$ can not be rejected. 
This result indicates that the effect of the new dam on the change of the amount of rainfall is small. }

{
\section{Further discussion} \label{sec8}
\def\theequation{7.\arabic{equation}}
\setcounter{equation}{0}

We conclude this paper with a brief discussion of the extension
of the proposed concept to the  multivariate case and its relation to the concept of sojourn times  in probability theory.

 \subsection{Multivariate data}

The results of this paper can be extended to multivariate  time series
of the form \begin{align}
\mathbf X_{i,n}=\boldsymbol \mu (i/n)+\mathbf{e} _{i,n}~,
\end{align}
where $\mathbf X_{i,n}=(X_{i,n}^1,...,X_{i,n}^m)^T$ is the $m$-dimensional vector of observations,
$\boldsymbol \mu(i/n)=(\mu^1(i/n),...,\mu^m(i/n))^T$ its corresponding expectation and  
$(\mathbf e_{i,n})_{i=1\ldots ,n} $ is an $m$-dimensional time series such that $\mathbf e_{i,n} = \boldsymbol G (i/n,\FF_i)$, where $\mathbf G(t,\FF_i)=(G_1(t,\FF_i),...,G_m(t,\FF_i))^T$ is an $m$-dimensional filter. Assume that the long run variance matrix 
$$
\Sigma(t) =\sum_{ {i}=-\infty}^\infty\mbox{cov} (\boldsymbol G(t,\FF_i),\boldsymbol G(t,\FF_0)) 
$$ of the 
error process is strictly positive and let $\|\mathbf v \|$  denote the Euclidean norm of an $m$-dimensional vector $\mathbf v$.
The excess mass for the $m$-dimensional mean function is then defined as
\begin{align}
\boldsymbol T_c:=\int_0^1\mathbf 1(\|\boldsymbol \mu(t)-\boldsymbol  \mu(0)\|>c)dt
\end{align}
and a test for the hypotheses  $H_0: \boldsymbol T_c\leq \Delta$ versus $H_1: \boldsymbol T_c>\Delta$ can be developed by estimating
this quantity  by 
\begin{align}
\hat{\boldsymbol T}_{N,c}=\frac{1}{N}\sum_{i=1}^N\int_{c^2}^\infty \frac{1}{h_d}K_d\left(\frac{ \|
 \hat{\boldsymbol\mu }
(i/N)-  \hat{\boldsymbol\mu } (0) \| ^2-u}{h_d}\right)du,
\end{align}
where $ \hat{\boldsymbol\mu } $ denote the vector of component-wise bias-corrected Jackknife 
estimates of the vector of regression functions.

The corresponding bootstrap test is now obtained by rejecting the null  hypothesis at level $\alpha$, whenever
\begin{align} \label{multtest}
nNb_nh_d \hat{\boldsymbol T}_{N,c}-\Delta>q_{1-\alpha}~, 
\end{align}
where $q_{1-\alpha}$ is the $(1-\alpha)$-quantile of the random variable
\begin{align}
\sum_{j=1}^n\sum_{i=1}^NK_d\left(\frac{\hat{\boldsymbol g}  (i/N)-c^2}{h_d}\right)\left( K^*\left(\frac{j/n-i/N}{b_n}\right)-\bar K^*\left(\frac{j}{nb_n}\right)\right)(\nabla \hat {\boldsymbol {g}}(i/N))^T \hat \Sigma^{1/2} (j/n)\boldsymbol V_j,
\end{align}
 $\nabla \hat{\boldsymbol g} (u)$ is the gradient of the function $
 \hat{\boldsymbol g} (u) =\| \hat{\boldsymbol \mu}  (u) -\hat{\boldsymbol \mu}  (0)\|^2 $, ${\boldsymbol V}_1 , {\boldsymbol V}_2, \ldots  $ are independent standard normally  distributed $m$-dimensional random vectors and  $\hat \Sigma(t)$ is an  analogue  of the long run variance matrix estimator defined in \eqref{2018-5.4}. 
 
 Under  similar conditions  as stated in in Assumption \ref{A1},  \ref{A2}, \ref{A4.1}, \ref{A4} and in  Theorem \ref{criticalversion}(a),
 an analogue of Theorem \ref{algorithm_5.1} can be proved, i.e. the 
 bootstrap test
defined by \eqref{multtest} has asymptotic level $\alpha$ and is consistent.

\subsection{Estimates of excess measures related to sojourn times}
\label{sec72}
{The excess measures \eqref{soj21} and \eqref{soj22}
based on sojourn times can easily be estimated under the assumption that 
 the process $\{ \epsilon (t) - \epsilon (0) \}_{t \in [0,1]} $ is stationary with density $f$. In this case the quantities $e_c$  and $p_{c,\Delta}$ can be expressed 
as
 \begin{align}
 \label{soj1}
    e_c&  =  \E(S_c)=\int \int_0^1 \mathbf 1 (| \mu(t) - \mu(0) +x | > c)f(x)dtdx, \\
     \label{soj2}
  p_{c,\Delta}&= \mathbb{P} (S_c > \Delta) =\E(\E(\mathbf 1(S_c>\Delta)| \epsilon(t)-\epsilon(0) =x)) \\
 &         =\int \mathbf 1\Big ( \int_{0}^1\mathbf 1( | \mu(t) - \mu (0) + x | > c)dt>\Delta \Big )f(x)dx, \notag
         \end{align}
        and corresponding estimators are given by  
\begin{align}
\hat e_c &=\frac{1}{Nnh_d}\sum_{i=1}^{n}\sum_{s=1}^N\int_c^\infty K_d\Big (\frac{|\hat \mu(s/N)-
\hat \mu(0)+\hat Z(i/n)|-u}{h_d}\Big )du,\\
\hat p_{c,\Delta}
&=\frac{1}{n}\sum_{i=1}^n\mathbf 1\Big (\frac{1}{Nh_d}\sum_{s=1}^N\int_c^\infty K_d\Big (\frac{|\hat \mu(s/N)-\hat \mu(0)+\hat Z(i/n)|-u}{h_d}\Big )du>\Delta\Big ),
\end{align}
respectively, where $\hat \mu(t)-\hat \mu(0)$ is a consistent estimator (say a local linear) of $\mu(t)-\mu(0)$ and 
$\hat Z (t) = \hat \epsilon(t)-\hat \epsilon(0)$ denotes  the corresponding residual.
Statistical analysis can then be developed along the lines of this paper. 
\\
However, in the case of a non-stationary error process as considered in this paper the situation is much more complicated and we leave the development of estimators and investigation of their 
(asymptotic) properties for future research.
}

\bigskip

\section*{Acknowledgements}
The authors would like to thank Martina Stein who typed this manuscript with considerable technical expertise and to V. Spokoiny for explaining his results to us and to V. Golosnoy for 
some help with the literature on control charts. The authors are also grateful to four unknown reviewers for their constructive comments on an earlier version of this manuscript.
The work of the authors was partially supported by
the Deutsche  Forschungsgemeinschaft (SFB 823: Statistik nichtlinearer dynamischer Prozesse, Teilprojekt A1 and C1, FOR 1735: Structural inference in statistics - adaptation and efficiency).


\bibliographystyle{imsart-nameyear}
\begin{small}
 \setlength{\bibsep}{2pt}
\bibliography{lit}
\end{small}

\bigskip 

\section{Proofs of main results}\label{sec7}
\def\theequation{8.\arabic{equation}}
\setcounter{equation}{0}

In this section we will prove the main results of this paper.
For the sake of a simple notation we
write $e_i:=\epsilon_{i,n}$ throughout this section, where $\epsilon_{i,n}$ is the non-stationary error process
in model \eqref{protype}. Moreover, in all
arguments given below $M$ denotes a sufficiently large constant which may vary  from line to line.
For the sake of brevity we will restrict ourselves to proofs of the results in Section \ref{sec4}, while the details for the proofs of the results in Section \ref{sec3} are omitted
as they follow by similar arguments as presented here. We will give a proof  of Theorem \ref{criticalversion} (deferring 
some of the more technical arguments to  the supplementary material) and  of Theorem \ref{algorithm_5.1} in this section.
The proof of Theorem  \ref{Thm6_new} can  also be found in  the supplementary material.

 \medskip

\subsection{ Proof of Theorem \ref{criticalversion}}
It follows from Assumption \ref{A1} that
there exist $k^+\geq 1 $ roots $t_1^+ < \ldots <t_{k^+}^+ $ of the equation $\mu(t)=\mu(0)+c$.
Define   $\gamma^+ = \min_{0\leq i\leq k^+}(t_{i+1}^+-t_i^+) >0$, with the  convention that $t_0^+=0$ and $t_{k^++1}^+=1$. 
Recalling the definition of the statistic $\tilde T_{N,c}^+$ and the quantity $ T_{N,c}^+$ in \eqref{statjack} and \eqref{tdet}, respectively, we obtain the decomposition
\begin{align} \label{basicdec}
\tilde  T_{N,c}^+-T_{N,c}^+=\Delta_{1,N}+\Delta_{2,N},
\end{align}
where the random variables
$\Delta_{1,N}$ and $\Delta_{1,N}$ are defined by
\begin{align}\label{delta12}
\begin{split}
\Delta_{1,N} &=\frac{1}{N}\sum_{i=1}^N\int_{c}^\infty\frac{1}{h_d^2}K_d'\Big(\frac{\mu(\frac{i}{N})-\mu(0)-u}{h_d}\Big)(\tilde \mu_{b_n}(\tfrac{i}{N})-\mu(\tfrac{i}{N})-(\tilde \mu_{b_n}(0)-\mu(0)) {)}du,\\
\Delta_{2,N} &=\frac{1}{2N}\sum_{i=1}^N\int_{c}^\infty\frac{1}{h_d^3}K_d''\Big(\frac{\zeta_i-u}{h_d}\Big)(\tilde \mu_{b_n}(\tfrac{i}{N})-\mu(\tfrac{i}{N})-(\tilde \mu_{b_n}(0)-\mu(0)) {)}^2du
\end{split}~~~~~~
\end{align}
(note that we do reflect the dependence of $\Delta_{\ell,N}$
on $n$ in our notation) and $\zeta_i$ denotes a random variable satisfying 
$| \zeta_i -(\mu(i/N)-\mu(0)) |
\leq  | \tilde \mu_{b_n}(i/N)-  \mu(i/N)-(\tilde \mu_{b_n}(0)-\tilde \mu_{}(0)) |$ and  $| \zeta_i -(\tilde \mu_{b_n}(i/N)-\tilde \mu_{b_n}(0)) |
\leq  | \tilde \mu_{b_n}(i/N)-  \mu(i/N)-(\tilde \mu_{b_n}(0)-\mu(0)) |$.
It is easy to see that
\begin{align}\label{Delta2}
|2\Delta_{2,N}|=\Big|\frac{1}{N}\sum_{i=1}^N\frac{1}{h_d^2}K_d'\Big(\frac{\zeta_i-c}{h_d}\Big)(\tilde \mu_{b_n}(i/N)-\mu(i/N)- {(}\tilde \mu_{b_n}(0)-\mu(0) {)})^2du\Big|.
\end{align}
Recall the definition of $\pi_n$ in \eqref{b1} and define 
 \begin{align}
 \label{SetAn}~~~~ A_n=\Big \{\sup_{t\in [b_n,1-b_n]\cup\{0\}}|\tilde \mu _{b_n} (t)-\mu(t)|\leq \pi_n, \sup_{t\in [0,b_n)\cup(1-b_n,b_n]}|\tilde \mu_{b_n} (t)-\mu(t)|\leq b_n^2\vee \pi_n \Big \},\end{align}
where we denote $\max\{a,b\}$ by $a\vee b$. By Lemma \ref{maxdevvar1} in Section \ref{sec8} of the online supplement, we have $\lim_{n\rightarrow \infty}\p(A_n)=1$ 
and Lemma \ref{sizeset} of the online supplement yields
\begin{align}
\sharp\{i:|\tilde \mu_{b_n} (i/N)-\tilde \mu_{b_n}(0)-c|\leq h_d, |\tilde \mu _{ b_n} (i/N)-\mu(i/N)-(\tilde \mu _{b_n} (0)-\mu(0))|\leq 2\pi_n\}
\notag\\\leq \sharp\{i:|\mu(i/N)-\mu(0)-c|\leq h_d+2\pi_n\}=O(N(h_d+\pi_n)^{1/(v^{+}+1)})\label{Appendix.6.17}
\end{align}
 almost surely,
where $\sharp A $  denotes the number of points in the set $A$.
Observing the definition of
$\zeta_i$ and \eqref{Appendix.6.17} we obtain
that the number of non-vanishing terms
on the right hand side of equality  \eqref{Delta2} is bounded by $O(N(h_d+\pi_n)^{\frac{1}{v+1}})$. Therefore the  triangle inequality yields for a sufficiently large constant
 $M$
\begin{align}
\|\Delta_{2,N}\mathbf 1(A_n)\|_2\leq M \Big(b_n^6+\frac{1}{nb_n}\Big)h^{-2}_d((h_d+\pi_n)^{\frac{1}{v^{+}+1}}).
\end{align}
Now Proposition B.3 of \cite{dette2015change}
 (note that $\lim_{n\rightarrow \infty}\p(A_n)=1$) yields the estimate
\begin{align}
\Delta_{2,N}=O_p\Big((b_n^6+\frac{1}{nb_n})h^{-2}_d((h_d+\pi_n)^{\frac{1}{v^{+}+1}}\Big).
\end{align}
Notice that the assumptions regarding   bandwidths guarantee that
\begin{align}\label{controldelta2}
\sqrt{nh_d}h_d^{\frac{v^+}{2(v^++1)}}\Delta_{2,N}&=o(1),\ \text{~~if  ~ $b_n^{v^++1}/h_d=r\in [0,\infty) $},\\
\sqrt{nb_n}h_d^{\frac{v^+}{v^++1}}\Delta_{2,N}& =o(1),\ \text{
~~if ~$b_n^{v^++1}/h_d\rightarrow \infty $}\label{controldelta2.1},
\end{align}
and therefore it remains to consider the term $\Delta_{1,N}$
in the decomposition \eqref{basicdec}.

For this purpose we recall its definition in \eqref{delta12}
and obtain by an application of
 Lemma  \ref{maxdevvar1} of the online supplement
and straightforward calculations
 the following decomposition
\begin{align} \label{IR}
\Delta^{}_{1,n}&=\frac{-1}{Nh_d}\sum_{i=1}^{N}K_d\Big(\frac{\mu(i/N)-\mu(0)-c}{h_d}\Big)\left(\left(\tilde \mu _{ b_n} (i/N)-\mu(i/N)\right)-\left(\tilde \mu_{ b_n} (0)-\mu(0)\right)\right)\notag\\&=I+R,
\end{align}
where the terms $I$ and $R$ are defined by
\begin{align}\label{new.42}
I &=\frac{-1}{nNb_nh_d}\sum_{i=1}^{N}K_d\Big(\frac{\mu(\tfrac{i}{N})-\mu(0)-c}{h_d}\Big)\sum_{j=1}^ne_j\Big(K^*\Big(\frac{\tfrac{i}{N}-\tfrac{j}{n}}{b_n}\Big)-\bar K^*\Big(\frac{j}{nb_n}\Big)\Big),\\
R&=O\Big(\frac{1}{Nh_d}\sum_{i=1}^NK_d\Big(\frac{\mu(\tfrac{i}{N})-\mu(0)-c}{h_d}\Big)\Big(b_n^3+\frac{1}{nb_n}\Big)\Big).
\end{align}
By Lemma \ref{sizeset} of the online supplement  the term $R$
is of order $O(h_d^{-\frac{v^+}{v^++1}}(b_n^3+\frac{1}{nb_n}))$.
For the investigation of the remaining term $I$, we use Proposition 5 of \cite{zhou2013}, which shows that  there exist (on a possibly richer probability space),
independent stand normally distributed random variables  $\{V_i\}_{i\in \mathbb Z}$, such that
\begin{align}
\max_{1\leq i\leq n}|\sum_{j=1}^ie_{j}-\sum_{j=1}^i\sigma(j/n)V_j|=o_p(n^{1/4}\log^2n).
\end{align}
This representation and the summation by parts formula in equation (44) of
 \cite{zhou2010nonparametric} yield
\begin{align}\label{Gaussian241}
\sup_{t\in [0,1]}
\Big |\sum_{j=1}^ne_j\tilde K^*\Big(\frac{t-j/n}{b_n}\Big)-\sum_{j= {1}}^i\sigma(j/n)V_j\tilde K^*\Big(\frac{t-j/n}{b_n}\Big) \Big |=o_p(n^{1/4}\log^2n),
\end{align}
where we  introduce the notation
\begin{align} \label{kstartilde}
 \tilde K^*\Big(\frac{t-j/n}{b_n}\Big)=K^*\Big(\frac{t-j/n}{b_n}\Big)-\bar K^*\Big(\frac{j}{nb_n}\Big).
\end{align}
Using these results in \eqref{new.42} and 
Lemma \ref{sizeset} of the online supplement provides an asymptotically equivalent representation of  the term $I$,  that is
\begin{align}\label{independent2}
|I'-I|=o_p\Big(\frac{n^{1/4}\log^2n}{nb_n}h_d^{-\frac{v^+}{v^++1}}\Big).
\end{align}
Here
\begin{align}
I':=\frac{-1}{nNb_nh_d}\sum_{j=1}^n\sum_{i=1}^{ N}K_d\Big(\frac{\mu(i/N)-\mu(0)-c}{h_d}\Big)\sigma(j/n)\tilde {K}^*\Big(\frac{i/N-j/n}{b_n}\Big)V_j\notag
\end{align}
is a zero mean Gaussian random variable with variance
{
\begin{align} \label{varIprime}
\mbox{Var}(I')&=\frac{1}{n^2b_n^2h_d^2}\sum_{j=1}^n\Big(\frac{1}{N}\sum_{i=1}^{N}\sigma(j/n)\tilde K^*\Big(\frac{i/N-j/n}{b_n}\Big)K_d\Big(\frac{\mu(i/N)-\mu(0)-c}{h_d}\Big)\Big)^2\notag
\\&=\frac{1}{n^2b_n^2h_d^2}\sum_{j=1}^n\Big(\int_{0}^{1}\sigma(j/n)\tilde K^*\Big(\frac{t-j/n}{b_n}\Big)K_d\Big(\frac{\mu(t)-\mu(0)-c}{h_d}
\Big)dt\Big)^2+ \beta_n\notag
\\&:=\bar \alpha_n+ \beta_n,
\end{align}
and the last two equalities define the quantities $\bar \alpha_n $ and $ \beta_n$ in an obvious manner.
Observing the estimates
\begin{align*}
&\frac{1}{N}\sum_{i=1}^N\sigma(\frac{j}{n})\tilde K^*\Big(\frac{i/N-j/n}{b_n}\Big)K_d\Big(\frac{\mu(i/N)-\mu(0)-c}{h_d}\Big)-\\
&\int_0^1\sigma(\frac{j}{n})\tilde K^*\Big(\frac{t-j/n}{b_n}\Big)K_d\Big(\frac{\mu(t)-\mu(0)-c}{h_d}\Big)dt
=O\Big((\frac{1}{Nb_n}+\frac{1}{Nh_d})(b_n\wedge h_d^{\frac{1}{v^++1}})\Big), \\
& \frac{1}{n^2b_n^2h_d^2}\sum_{j=1}^n\Big(\int_{0}^{1}\sigma(j/n)\tilde K^*\Big(\frac{t-j/n}{b_n}\Big)K_d\Big(\frac{\mu(t)-\mu(0)-c}{h_d}\Big)dt\Big)
=O\Big(\frac{h_d^{\frac{-v^+}{v^++1}}}{nb_nh_d}\Big),
\end{align*}
we have that
\begin{align}\label{new.55-June}
\beta_n=\frac{h_d^{\frac{-v^+}{v^++1}}}{nb_nh_d}\Big(\frac{1}{Nb_n}+\frac{1}{Nh_d}\Big)(b_n\wedge h_d^{\frac{1}{v^++1}})
+\Big((\frac{1}{Nb_n}+\frac{1}{Nh_d})(b_n\wedge h_d^{\frac{1}{v^++1}})\Big)^2,
\end{align}
where $a\wedge b:=\min(a,b)$.}

For the  calculation of  $\bar \alpha_n$ we  note that
\begin{align} 
\label{eqstar} \bar{K}^*\Big(\frac{j/n}{b_n}\Big)K^*\Big(\frac{t-j/n}{b_n}\Big)K_d\Big(\frac{\mu(t)-\mu(0)-c}{h_d}
\Big)=0.
\end{align}
for   sufficiently large $n$. This statement follows because by Lemma \ref{sizeset} of the online supplement  the third factor vanishes outside of (shrinking)  neighbourhoods  ${\cal U}_1 , \ldots ,{\cal U}_{k^+}$  
of the points $t_1^+ , \ldots , t_{k^+}^+$ with Lebesgue measure of order $h_d^{\frac{1}{v_l+1}}$, ($1\leq l\leq k^+$).
Consequently, the product of the first and second factor vanishes, wheneever the point 
$j/n$ is {\bf not} an element of the set 
$$
 \big \{  s + t  ~\big | ~ t \in \cup_{j=1}^{k^+}{\cal U}_j ~ ; s  \in [-b_n,b_n]  \big \} .
$$
However, if  $n$ is sufficiently large the intersection of this set with the interval $[0,b_n]$, is empty. Consequently, 
for sufficiently large $n$  there exists no pair $(t, j/n)$ such that all factors in \eqref{eqstar} different from zero. \\
Therefore,  we obtain (recalling  the notation of $\tilde K^*$ in  \eqref{kstartilde})
\begin{align} \label{alphazerl}
\bar \alpha_n
=\alpha_n+\tilde {\alpha}_n,
\end{align}
where
\begin{align}\label{new.51}
\alpha_n &=\frac{1}{n^2b_n^2h_d^2}\sum_{j=1}^n\Big(\int_{0}^{1}\sigma(j/n) K^*\Big(\frac{t-j/n}{b_n}\Big)K_d\Big(\frac{\mu(t)-\mu(0)-c}{h_d}
\Big)dt\Big)^2,\\
\tilde \alpha_n
& =\frac{1}{n^2b_n^2h_d^2}\sum_{j=1}^n\Big(\int_{0}^{1}\sigma(j/n)\bar K^*\Big(\frac{j/n}{b_n}\Big)K_d\Big(\frac{\mu(t)-\mu(0)-c}{h_d}
\Big)dt\Big)^2.\label{new.52}
\end{align}
{
In the supplementary material we will show   that
\begin{align}\label{an}
\alpha_n &=
\left\{
\begin{array}{cl}
{h_d^{\frac{-2v^+}{v^++1}}}({nb_n})^{-1} \sigma_1^{2,+} & \mbox{ if }  b^{v^{+}+1}_n/h_d \rightarrow \infty  \\
& \\{h_d^{\frac{1}{v^++1}}}({nh_d^2})^{-1} \rho_1^{2,+}
& \mbox{ if }
b^{v^++1}_n/h_d \rightarrow r  \in [0,\infty)
\end{array}
\right.
\end{align}
\begin{align}
\tilde \alpha_n
& =
\left\{
\begin{array}{cl}
{h_d^{\frac{-2v^+}{v^++1}}}({nb_n})^{-1} \sigma_2^{2,+} & \mbox{ if }  b^{v^++1}_n/h_d \rightarrow \infty  \\
& \\{h_d^{\frac{1}{v^++1}}}({nh_d^2})^{-1} \rho_2^{2,+}
& \mbox{ if }
b^{v^++1}_n/h_d \rightarrow r \in [0,\infty)
\end{array}
\right.
~.\label{anbar}
\end{align}
where $\sigma_1^{2,+} $, $\sigma_2^{2,+} $,
$\rho_1^{2,+} $  and $\rho_2^{2,+} $ are defined in Theorem
\ref{criticalversion}.
The assertion now follows from \eqref{basicdec}, \eqref{controldelta2}, \eqref{controldelta2.1},
\eqref{IR} and \eqref{independent2}
observing that the random variable $I^\prime $ is normally distributed, where the (asymptotic) variance can be obtained from \eqref{varIprime}, \eqref{new.55-June}, \eqref{alphazerl}, \eqref{an} and \eqref{anbar}.
} 
$\Box$

\bigskip

\subsection{ Proof of Theorem \ref{algorithm_5.1} }
We have to distinguish two cases:

\smallskip
\noindent
{\bf (1) The equation $\mu(t)-\mu(0)=c$ has at least one solution.}
Recall the definition of  the quantity $I'$ in   \eqref{iprime}, then
it follows from the proof of Theorem \ref{criticalversion},  that
\begin{align}
\mbox{Var}(\sqrt{nb_n}h_d^{\frac{v^+}{v^++1}}I')=\sigma^{{2,+}}_1+\sigma^{{2,+}}_2+o(1),
\end{align}
where $\sigma^{{2,+}}_1$ and  $\sigma^{{2,+}}_2$
are defined in \eqref{DefinitionU4a} and \eqref{DefinitionU4}, respectively.
Note that $\mbox{Var}(I')=\frac{1}{n^2N^2b^2_nh^2_d}\tilde V$, where
\begin{align*}
\tilde V=\sum_{j=1}^n\sigma^2(j/{n})\Big (
\sum_{i=1}^NK_d\Big (\frac{ \mu_{}(i/N)- \mu_{}(0)-c}{h_d}\Big )\Big (K^*\Big (\frac{i/N-j/n}{b_n}\Big)-\bar K^*\Big(\frac{j}{nb_n}\Big)\Big)\Big )^2.
\end{align*}
At the end of this proof we will show that
\begin{align}\label{barV}
\frac{(\sqrt{nb_n}h_d^{\frac{v^+}{v^++1}})^2}{n^2N^2b^2_nh^2_d}(\tilde V-\bar V)=o(1),
\end{align}
which implies that
\begin{align} \label{quantcon}
\lim_{n\to \infty} {\sqrt{nb_n}h_d^\frac{v^+}{v^++1}}q^+_{1-\alpha}/({nNb_nh_d})
=\Phi^{-1}(1-\alpha)\sqrt{\sigma_1^{{2,+}}+\sigma_2^{{2,+}}}.
\end{align}
Observing the identity
\begin{align}
\label{new.5.6}
&\mathbb{P} \big ( nNb_nh_d\big (\tilde T_{N,c}^+-\Delta \big )>q^+_{1-\alpha} \big ) \\&=\mathbb{P} \left ( \frac{\sqrt{nb_n}h_d^{\frac{v^+}{v^++1}}\big(\tilde T_{N,c}^+-T_c^+\big )}{\sqrt{\sigma_1^{{2,+}}+\sigma_2^{{2,+}}}}> \frac{\frac{\sqrt{nb_n}h_d^{\frac{v^+}{v^++1}}}{ nNb_nh_d}q^{+}_{1-\alpha}+\sqrt{nb_n}h_d^\frac{v^+}{v^++1}(\Delta-T_c^+)}{\sqrt{\sigma_1^{{2,+}}+\sigma_2^{{2,+}}}} \right)~\notag
\end{align}
the assertion now follows from \eqref{quantcon} and Theorem
\ref{criticalversion}, which shows that the random variable
$$\frac{\sqrt{nb_n}h_d^{\frac{v^+}{v^++1}}\big(\tilde T_{N,c}^+-T_c^+ \big )}{\sqrt{\sigma_1^{{2,+}}+\sigma_2^{{2,+}}}}$$
converges weakly to a standard normal distribution.

\noindent
It remains to prove  \eqref{barV}, which is a consequence of the following observations

\begin{itemize}
	\item[(a)]
	$\hat \sigma(t^+_l)=\sigma(t^+_l)(1+o(1))$, uniformly with respect to $l=1, \ldots , k^+$.

	\item[(b)]  The bandwidth condition $\pi_n/h_d=o(1)$,
	Proposition \ref{prop1}
	and similar arguments as \eqref{Appendix.6.17} show
	$$
	K_d\Big(\tfrac{\mu(\frac{i}{N})-\mu(0)-c}{h_d}\Big)-K_d\Big(\tfrac{\tilde \mu_{b_n}(\frac{i}{N})-\tilde \mu_{b_n} (0)-c}{h_d}\Big)=O\Big (\sum_{ {\{l:v_l^+=v^+\}}} \mathbf 1\big (
	\big |\tfrac{i}{N}-t^+_l \big |\leq h_d^{\tfrac{1}{v^++1}}\big)\tfrac{\pi_n}{h_d} \Big),
	$$
	where $\pi_n$ is defined in Theorem \ref{criticalversion}.
\end{itemize}
This completes the proof  of Theorem \ref{algorithm_5.1} in the case that there exist in fact roots of the equation $ \mu (t)- \mu (0) =c $.

\medskip
\noindent
{\bf  (2) The  equation $\mu(t)-\mu(0)=c$ has no solutions. }
In this case we have  $\mu(t)-\mu(0)<c$ where $c>0$. 
Note that for two sequences of measurable sets $U_n$ and $V_n$ such that $\p(U_n)\rightarrow 1$ and $\p(U_n\cap V_n)\rightarrow u\in (0,1)$, we have
$\p(V_n)\rightarrow u$.
Consequently, as the set $A_n$ defined in \eqref{SetAn} satisfies  $\p(A_n)\rightarrow 1$ the assertion of 
the theorem follows from 
\begin{align}
\lim_{n\rightarrow \infty}\p(nNb_nh_d(\tilde T_{N,c}^+-\Delta)>q_{1-\alpha}^+,A_n,\mu(t)-\mu(0)<c)=0.\label{2018-5.34}
\end{align}
However, under the event $A_n$ and $\mu(t)-\mu(0)<c$ we have $q_{1-\alpha}^+=0$ and $\tilde T_{N,c}^+=0$, if  $n$ is sufficiently large. Thus 
\eqref{2018-5.34} is obvious (note that $0<\Delta<1$), which finishes the proof in the case where   the equation $ \mu (t)- \mu (0) =c$ has in fact no roots.
\hfill $\Box$

\bigskip

{\centering{\bf Appendix}}\ \\\ \\
In this section we will provide technical details for the proof of Theorem \ref{criticalversion} and a
proof of Theorem \ref{Thm6_new}. Recall that we use the notation 
$e_i:=\epsilon_{i,n}$ throughout this section, where $\epsilon_{i,n}$ is the nonstationary error process
in model \eqref{protype}. Moreover, in all
arguments given below $M$ denotes a sufficiently large constant which may vary  from line to line.

 \section{Proof of of Theorem \ref{criticalversion} and  \ref{Thm6_new}}

\noindent
{\bf Proof of Theorem \ref{criticalversion}.} Following the arguments of the main article, it remains to show \eqref{an} and \eqref{anbar} to complete the proof of Theorem \ref{criticalversion}.

\smallskip\noindent
{\it Proof of \eqref{an}}:
By Lemma \ref{sizeset} with $m$ replaced by $\mu(0)+c$, there exists a small positive number $0<\epsilon<\gamma^{+}/4$ such that when $n$ is sufficiently large, we have
\begin{align}\label{newnewsquare60}
\alpha_n&=\frac{1}{n^2b_n^2h_d^2}\sum_{j=1}^n\Big(\sum_{l=1}^{k^{+}}\int_{t_l^{+}-\epsilon}^{t_l^{+}+\epsilon}\sigma(j/n)K^*\Big(\frac{t-j/n}{b_n}\Big)K_d\Big(\frac{\mu(t)-\mu(0)-c}{h_d}
\Big)dt\Big)^2\notag
\\&=\frac{1}{n^2b_n^2h_d^2}\sum_{j=1}^n\sum_{l=1}^{k^{+}}\Big(\int_{t_l^{+}-\epsilon}^{t_l^{+}+\epsilon}\sigma(j/n)K^*\Big(\frac{t-j/n}{b_n}\Big)K_d\Big(\frac{\mu(t)-\mu(0)-c}{h_d}
\Big)dt\Big)^2\notag
\\&=\frac{1}{n^2b_n^2h_d^2}\sum_{j=1}^n\sum_{l=1}^{k^{+}}\alpha^2_{n,l,j},
\end{align}
where the last equation defines the quantities $\alpha_{n,l,j}^2$ in an obvious manner. 
We now calculate $\alpha_n$ for the  two bandwidth conditions in \eqref{an}.
\medskip

\noindent (i) We begin with the case  $b^{v^++1}_n/h_d \rightarrow \infty $, which means $b^{v^+_l+1}_n/h_d \rightarrow \infty $ for $l=1,\dots,k$.
By Lemma \ref{sizeset}  there exists a sufficiently large constant $M$ such that
\begin{align}
\ \ \ \ \ \ \alpha_{n,l,j}=
\int_{t_l^{+}-Mh_d^{\frac{1}{v^+_l+1}}}^{t_l^{+}+Mh_d^{\frac{1}{v^+_l+1}}}\sigma(t_l^{+})K^*\Big(\frac{t-j/n}{b_n}\Big)K_d
\Big(\frac{\mu(t)-\mu(0)-c}{h_d}\Big)dt\Big(1+O\Big(h_d^{\frac{1}{v^+_l+1}}\Big)\Big).\label{new.79-June}
\end{align}
Observing the fact that the kernel $K_d(\cdot)$ is bounded and continuous we obtain by a
Taylor expansion of $\mu(t)-\mu(0)-c$ around $t_l^{+}$, 
{
	\begin{align}
	\big|\alpha_{n,l,j}   -  \alpha_{n,l,j}^* \big|
	=O\Big(h_d^{\frac{2}{v^+_l+1}}\mathbf
	1(|j/n-t^+_l|\leq 2b_n)\Big) \label{argu65}
	\end{align}
	uniformly with respect to $t\in [0,1]$,
	where
	\begin{align*}
	\alpha_{n,l,j}^* =  \int_{t_l^{+}-Mh_d^{\frac{1}{v^+_l+1}}}^{t_l^{+}+Mh_d^{\frac{1}{v^+_l+1}}}\sigma(t_l^{+})K^*\Big(\frac{t-j/n}{b_n}\Big)K_d\Big(\frac{\mu^{(v^+_l+1)}(t^{+}_l)(t-t^+_l)^{v^+_l+1}}{(v^+_l+1)!h_d}\Big)dt.
	\end{align*}
	Substituting $t=t_l^{+}+z\big |{h_d(v^+_l+1)!}/ {\mu^{(v^+_l+1)}(t_l^+)}\big |^{\frac{1}{v^+_l+1}}$, observing   the symmetry of $K_d(\cdot)$ and using a Taylor expansion shows that
	\begin{align}\label{square63}
	\alpha_{n,l,j}^*
	&=\Big|\frac{(v^+_l+1)!h_d}{\mu^{(v^+_l+1)}(t^{+}_l)}\Big|^{\frac{1}{v^+_l+1}}\Big(\int K_d(z^{v^+_l+1})dz\Big)\sigma(t^{+}_l)K^*\Big(\frac{t^{+}_l-j/n}{b_n}\Big)
	\\
	\nonumber
	& ~~~~~~~ +O\Big ({h_d^{\frac{2}{v^+_l+1}}}({b_n})^{-1}
	\mathbf 1(|j/n-t^{+}_l|\leq 2b_n)\Big ),
	\end{align}
	where we have used the fact that $\int zK_d(z^{v^+_l+1})dz< \infty$ since $K_d(\cdot)$ has a compact support.
	Equations 
	\eqref{new.79-June}--\eqref{square63} and the condition  $\frac{b_n^{v^++1}}{h_d}\rightarrow \infty$
	now give
	\begin{align}
	\alpha_n&=\frac{1}{n^2b_n^2h_d^2}\sum_{l=1}^k\sum_{j=1}^n\Big(\Big(\int K_d(z^{v_l^++1})dz\Big)\sigma(t^{+}_l)
	\Big|\frac{(v^++1)!h_d}{\mu^{(v^++1)}(t_l^+)}\Big|^{\frac{1}{v^++1}}K^*\Big(\frac{t_l^+-j/n}{b_n}\Big)\Big)^2\notag\\&
	~~~~~~~~~~~~~~~\times
	\left (1+O\big ({h_d^{\frac{1}{v^++1}}}{b_n}^{-1} \mathbf 1(|j/n-t^{+}_l|\leq 2b_n)\big )\right)\notag
	\\&=\sum_{l=1}^k\frac{h_d^{\frac{-2v_l^+}{v_l^++1}}}{nb_n}\Big(\int K_d(z^{v_l^++1})dz\Big)^2((v_l^++1)!)^{\frac{2}{v_l^++1}}\Big(\frac{\sigma(t_l^+)}{|\mu^{(v_l^++1)}(t_l^{+})|^{\frac{1}{v^{+}_l+1}}}\Big)^2
	\int (K^*(x))^2dx\notag\\
	& ~~~~~~~~~~~~~~~\times
	\big (1+O\big ( {(}nb_n {)}^{-1} +h_d^{\frac{1}{v^+_l+1}}/b_n\big )\big ) \nonumber  \\
	&= {h_d^{\frac{-2v^+}{v^++1}}}({nb_n})^{-1} \sigma_1^{2,+}
	\big ( 1 + o(1) \big )  \nonumber
	\end{align}
	which proves \eqref{an} in the case $b^{v^++1}_n/h_d \rightarrow \infty$.
}

\medskip
\noindent
Next we turn to the case   $b_n/h_d^{\frac{1}{v^++1}}\rightarrow c\in [0,\infty) $, introduce the notation
$\alpha_{n,l}=\frac{1}{n^2b_n^2h_d^2}\sum_{j=1}^n\alpha^2_{n,l,j}$
and note that
\begin{align} \label{decomp1}
\alpha_n = \sum_{l=1}^k \alpha_{n,l}.
\end{align}
Define $c_l=b_n^{v^+_l+1}/h_d $ 
for $l\in\{1,\ldots ,k^+\}$. For those $l$ satisfying $c_l\rightarrow \infty$, we have already  shown  that
\begin{align}
\alpha_{n,l}=\frac{h_d^{\frac{-2v^+_l}{v^+_l+1}}}{nb_n}= o\Big(\frac{h_d^{\frac{1}{v^+_l+1}}}{nh_d^2}\Big)=o\Big(\frac{h_d^{\frac{1}{v^++1}}}{nh_d^2}\Big).
\end{align}
In the following discussion  we prove that for those
$l$, for which $c_l$ does not converge to infinity,
the quantity  $\alpha_{n,l}$ is exactly of order $O({h_d^{\frac{1}{v^+_l+1}}}({nh_d^2})^{-1})$.
{
	For this purpose define
	\begin{align} \label{alphaprime}
	\alpha'_{n,l}=\frac{1}{nb_n^2h_d^2}\int_{0}^1
	\big ( G(t^{+}_l,s,b_n,h_d) \big )^2ds.
	\end{align}
	where
	$$
	G(t^{+}_l,s,b_n,h_d)=\int_{t^{+}_l-\epsilon}^{t_l^{+}+\epsilon}\sigma(s)K^*\Big(\frac{t-s}{b_n}\Big)K_d\Big(\frac{\mu(t)-\mu(0)-c}{h_d}\Big)dt.
	$$
	It follows from a Taylor expansion and an approximation by a  Riemann sum that
	\begin{align}\label{argu50}
	|\alpha_{n,l}-\alpha_{n,l}'|\leq \frac{1}{nb_n^2h_d^2}\sum_{j=1}^n\frac{1}{n^2}\sup_{\frac{j-1}{n}\leq s\leq \frac{j}{n}}
	\left|G(t^{+}_l,s,b_n,h_d)\right|\Big|\frac{\partial}{\partial s}G(t^{+}_l,s,b_n,h_d)\Big|.
	\end{align}
	The terms in this sum can be estimated by an application of Lemma \ref{sizeset}, that is
	\begin{align}
	\label{new.71}
	\sup_{\frac{j-1}{n}<s\leq \frac{j}{n}}
	\left|G(t^{+}_l,s,b_n,h_d)\right|
	& \leq C\lambda\left( {\cal D}_{lj} \right),\\
	\sup_{\frac{j-1}{n}<s\leq \frac{j}{n}}\Big|\frac{\partial}{\partial s}G_j(t^{+}_l,s,b,h_d)\Big|
	& \leq C\lambda\left( {\cal D}_{lj}  \right)/b_n,\label{new.72}
	\end{align}
	where
	$$
	{\cal D}_{lj} =
	\Big (\frac{j-1}{n}-b_n,\frac{j+1}{n}+b_n \Big )\cap
	\Big (t^+_l-Mh_d^{\frac{1}{v_l^++1}},t^+_l+Mh_d^{\frac{1}{v_l^++1}} \Big )~,
	$$
	$M$ and $C$ are sufficiently large constants and
	$\lambda(\cdot)$ denotes  the Lebesgue measure.
	Straightforward calculations show that the number of
	indices $j$ such that the set ${\cal D}_{lj} $ is not empty
	is of order  $O(nh_d^{\frac{1}{v^+_l+1}})$, while the Lebesgue measure in  \eqref{new.71} and of \eqref{new.72} is of order
	$O(b_n)$ and $O(1)$, respectively.
}
Combining these facts we obtain
\begin{align}\label{argu73}
\alpha_{n,l}=\alpha_{n,l}'+O\Big(nh_d^{\frac{1}{v^+_l+1}}b_n\frac{1}{n^3b_n^2h_d^2}\Big)
\end{align}
(for all $l=1, \ldots , k^+$ such that $c_l<\infty$). 
As the function  $\sigma $ is strictly positive on a compact set it follows that
\begin{align}\label{argu74}\alpha_{n,l}'=\alpha_{n,l}''\Big(1+O\Big(b_n+h_{d}^{\frac{1}{v^+_l+1}}\Big)\Big)
,\end{align}
where the quantity $\alpha_{n,l}''$ is defined as $\alpha_{n,l}'$ in \eqref{alphaprime}
replacing the $\sigma(s)$ by  $\sigma(t^+_l)$.  Define
\begin{align}\label{new.64}
\alpha_{n,l}'''=\frac{\sigma^2(t^+_l)}{nb_n^2h_d^2}\int_{0}^1\Big(\int_{t^+_l-\epsilon}^{t^+_l+\epsilon}K^*\Big(\frac{t-s}{b_n}\Big)K_d\Big(\frac{\mu^{(v^+_l+1)}(t^+_l)(t-t^+_l)^{v^+_l+1}}{(v^+_l+1)!h_d}
\Big)dt\Big)^2ds
\end{align}
and note that the only difference between $\alpha_{n,l}''$ and $\alpha_{n,l}'''$ is the term inside $K_d(\cdot)$.
{ A Taylor expansion around $t^+_l$ yields 
	$$
	\frac{\mu(t)-\mu(0)-c}{h_d}=\frac{\mu^{(v_l^++1)}(t_l^*)(t-t_l^+)^{v_l^++1}}{(v_l^++1)!h_d}
	$$
	for  some $t_l^*\in [t_l\wedge t^*_l,t_l\vee t^*_l]$
	and the mean value theorem gives
	\begin{align}
	&K_d\Big(\frac{\mu(t)-\mu(c)-c}{h_d}\Big)-K_d\Big(\frac{\mu^{(v^+_l+1)}(t^+_l)(t-t^+_l)^{v^+_l+1}}{(v^+_l+1)!h_d}\Big)
	\notag\\&=K_d'\Big( \frac{((1-\theta_l)\mu^{(v_l^++1)}(t_l)+\theta_l\mu^{(v^+_l+1)}(t_l^*))(t-t^+_l)^{v^+_l+1}}{(v^+_l+1)!h_d}\Big)\frac{(\mu^{(v^+_l+1)}(t_l^*)-\mu^{(v^+_l+1)}(t_l))(t-t^+_l)^{v^+_l+1}}{(v^+_l+1)!h_d}
	\end{align}
} 
for some $\theta_l\in [-1,1]$. 
Then  similar arguments as used in the derivation
of \eqref{argu73} show that
\begin{align}\label{argu77}
\alpha_{n,l}'''-\alpha_{n,l}''=O\Big({h_d^{-\frac{2v^+_l}{v^+_l+1}}}{n}^{-1}\Big).
\end{align}
On the other hand, further expanding the squared term of \eqref{new.64} yields that
\begin{align}\label{new.78}
\alpha'''_{n,l}=\frac{\sigma^2(t^+_l)}{nb_n^2h_d^2}\int_{0}^1\int_{t^+_l-\epsilon}^{t^+_l+\epsilon}\int_{t^+_l-\epsilon}^{t^+_l+\epsilon}K^*\Big(\frac{t-s}{b_n}\Big)K_d\Big(\frac{\mu^{(v^+_l+1)}(t^+_l)(t-t^+_l)^{v^+_l+1}}{(v^+_l+1)!h_d}
\Big)\notag\\
\times K^*\Big(\frac{v-s}{b_n}\Big)K_d\Big(\frac{\mu^{(v^+_l+1)}(t^+_l)( {v}-t^+_l)^{v^+_l+1}}{(v^+_l+1)!h_d}
\Big)dvdt ds.
\end{align}
For $t,v$ satisfying $|t-t^+_l|=O(\min\{b_n,h_d^{\frac{1}{v^+_l+1}}\})$,
$|v-t_l|=O(\min\{b_n,h_d^{\frac{1}{v^+_l+1}}\})$, straightforward calculations show
\begin{align}\label{new.4.80}
\int_0^1K^*\Big(\frac{t-s}{b_n}\Big)K^*\Big(\frac{v-s}{b_n}\Big)ds=b_n\int_{-\infty}^\infty
K^*\Big(u\Big)K^*\Big(\frac{v-t}{b_n}+u\Big)du .
\end{align}
To move forward, we introduce the notation
\begin{align*}
z_1 =
(t-t^{+}_l)
\Big|\tfrac{\mu^{(v^+_l+1)}(t^+_l)}{h_d(v^+_l+1)!}\Big|^{\frac{1}{v^+_l+1}},~z_2=
(v-t^{+}_l)
\Big|\frac{\mu^{(v^+_l+1)}(t^+_l)}{h_d(v^+_l+1)!}\Big|^{\tfrac{1}{v^+_l+1}},~
\theta(v^+_l,h_d)=\Big|\tfrac{h_d(v^+_l+1)!}{\mu^{(v^+_l+1)}(t^+_l)}\Big|^{\frac{1}{v^+_l+1}}~.
\end{align*}
By a change of variables and \eqref{new.4.80},
we now obtain
\begin{align}
\alpha_{n,l}'''&=\frac{\sigma^2(t_l^{+})\theta(v^+_l,h_d)^2}{nb_nh_d^2}\int\int\int K^*(u)K^*\Big(u+\frac{1}{b_n}\theta(v^+_l,h_d)(z_2-z_1)\Big)\notag\\
& ~~\times
K_d(z_1^{v^+_l+1})K_d(z_2^{v^+_l+1})dz_1dz_2du\notag\\
=&\frac{\sigma^2(t^+_l)}{nh_d^2}h_d^{\frac{1}{v^+_l+1}}\Big|\frac{(v^+_l+1)!}{\mu^{(v^+_l+1)}(t^+_l)}\Big|^{\frac{1}{v^+_l+1}}
\int\int\int K^*(u)K^*(v)K_d(z_1^{v^+_l+1})\notag
\\&
~~\times
K_d\Big (\Big(z_1+c_l\Big |\frac{(v^+_l+1)!}{\mu^{(v^+_l+1)}(t^+_l)}\Big|^{\frac{-1}{v^+_l+1}}(v-u)\Big)
^{v_l^++1}\Big )dudvdz_1
\end{align}
{
	Finally, combining  \eqref{argu73}, \eqref{argu74} and \eqref{argu77} we have that
	\begin{align}
	\alpha_{n,l}=\alpha_{n,l}'''\Big(1+b_n+\frac{1}{nb_n}+h_d^\frac{1}{v^+_l+1}\Big)~,
	\end{align}
	and, observing that $h_d^{\frac{1}{v^+_l+1}}=o(h_d^{\frac{1}{v^++1}})$
	whenever $v^+_l < v^+$,
	we obtain from \eqref{decomp1}
	\begin{align*}
	\alpha_n &=\frac{|h_d(v^++1)!|^{\frac{1}{v^++1}}}{nh_d^2}\sum_{\{l:v^+_l=v^+\}}\frac{\sigma^2(t^+_l)}{|\mu^{(v^++1)}(t^+_l)|^{\frac{1}{v^++1}}}
	\int\int\int K^*(u)K^*(v)K_d(z_1^{v^++1})\notag
	\\&\times K_d\Big(\Big(z_1+r\Big|\frac{(v^++1)!}{\mu^{(v^++1)}(t^+_l)}\Big|^{\frac{-1}{v^++1}}(v-u)\Big)
	^{v^++1}\Big)dudvdz_1(1+o(1))~,
	\end{align*}
	which proves \eqref{an} in the case $b^{v^++1}_n/h_d \rightarrow r\in[0,\infty)$.
}

\medskip
\noindent
{\it Proof of \eqref{anbar}.}
Recalling  the definition of $\tilde \alpha_n$ in \eqref{new.52} we obtain by straightforward calculations
and  a Taylor expansion
\begin{align}
\tilde \alpha_{n}=\frac{\sigma^2(0)}{nb_nh_d^2}\int_0^1 (\bar K^*(t))^2dt\Big(\int_0^1K_d\Big(\frac{\mu(t)-\mu(0)-c}{h_d}\Big)dt\Big)^2\Big(1+O\Big(b_n+\frac{1}{nb_n}\Big)\Big).\label{Nov27-5.49}
\end{align}
Similar (but easier) arguments as used in the derivation of \eqref{argu65} and \eqref{square63} show
\begin{align}
\int_0^1K_d\Big(\frac{\mu(t)-\mu(0)-c}{h_d}\Big)dt=&|h_d(v^++1)!|^{\frac{1}{v^++1}}\notag\\ 
\times &\sum_{\{l:v^+_l=v^+\}}|\mu^{(v^++1)}(t^+_l)|^{-\frac{1}{v^++1}}
\int K_d(z^{v^++1})dz(1+o(1)),\label{Nov27-5.48}
\end{align}
which gives
\begin{align*}
\tilde \alpha_n &=\frac{\sigma^2(0)h_d^{\frac{-2v^+}{v^++1}}((v^++1)!)^{\frac{2}{v^++1}}}{nb_n}\int_0^1 (\bar K^*(t))^2dt
\\&\times \Big(\sum_{\{l:v^+_l=v^+\}}|\mu^{(v^++1)}(t^+_l)|^{-\frac{1}{v^++1}}
\int K_d(z^{v^++1})dz\Big)^2(1+o(1)).
\end{align*}
Consequently, if $ b^{v^++1}_n/h_d \rightarrow \infty $
we  have
\begin{align*}
\tilde \alpha_n=\frac{h_d^{\frac{-2v^+}{v^++1}}}{nb_n}
\sigma_2^{2,+} (1+o(1)).
\end{align*}
where $\sigma_2^{2,+}$ is defined by \eqref{DefinitionU4}.
This proves the statement \eqref{anbar} in the case
$ b^{v^++1}_n/h_d \rightarrow \infty $, while the second case follows by similar arguments observing
that  we have
$$
nh_d^{\frac{v^+}{v^++1}+1}\frac{h_d^{-\frac{2v^+}{v^++1}}}{nb_n}=r^{-1}
$$
if $ b^{v^++1}_n/h_d \rightarrow r^{v^++1} \in [0,\infty) $.
\noindent

\bigskip

\noindent
{\bf Proof of Theorem \ref{Thm6_new}.} Define
$\tilde S_{k,r}=\sum_{i=k\vee 1}^{r\wedge n}e_{i}$,
$$
\tilde \Delta_j=\frac{\tilde S_{j-m+1,j}-\tilde S_{j+1,j+m}}{m}
~,~~
\tilde{\sigma}^2(t)=\sum_{j=1}^n\frac{m\tilde \Delta_j^2}{2}w(t,j).$$ 
Since $\mu(\cdot)\in \mathcal C^2$, elementary calculations show that uniformly for $t\in [0,1]$,
\begin{align}
|\tilde \sigma^2(t)-\hat \sigma^2(t)|=O_p(m^{5/2}/n)\label{2018-6.6}
\end{align} 
Similar arguments as given in the proof of
Lemma 3 of \cite{zhou2010simultaneous} yields 
$\sup_j\|\tilde \Delta_j\|_4=O(m^{-1/2})$.
A further application of this lemma  gives
\begin{align}\label{2018-6.6}
\|\sup_{t\in [\gamma_n,1-\gamma_n]}|\tilde \sigma^2(t)-\E(\tilde \sigma^2(t))|\|_2=O(m^{1/2}n^{-1/2}\tau_n^{-1}),\\
\|\tilde \sigma^2(t)-\E(\tilde \sigma^2(t))\|_2=O(m^{1/2}n^{-1/2}\tau_n^{-1/2})\label{2018-6.7}
\end{align}
Elementary calculations show that 
\begin{align}\label{2018-6.8}
\E(\tilde \sigma^2(t))=\Lambda_1(t)+\Lambda_2(t)+\Lambda_3(t),
\end{align}
where
\begin{align*}
\Lambda_1(t)
&=\frac{1}{2m}\sum_{j=1}^n\tilde{S}_{j-m-1,j}^2\omega(t,j), \\
\Lambda_2(t)
&=\frac{1}{2m}\sum_{j=1}^n\tilde{S}_{j+1,j+m}^2\omega(t,j), \\
\Lambda_3(t)
&=-\frac{1}{2m}\sum_{j=1}^n \tilde{S}_{j+1,j+m}\tilde{S}_{j-m-1,j}\omega(t,j).
\end{align*}
Recall the representation $e_i=G(i/n,\FF_i)$. Define $\tilde S^\diamond_{j-m+1,j}=\sum_{r=1 \vee (j-m+1)}^jG(j/n,\FF_{r})$, and $\tilde S^\diamond_{j-m+1}=\sum_{r=j+1}^{n\wedge(j+m)}G(j/n,\FF_{r})$. For $s=1,2,3$, define $\Lambda^\diamond_s(t)$ as the quantity where the terms $\tilde S_{j-m+1,j}$ and $\tilde S_{j-m+1}$ in $\Lambda_s(t)$ are replaced by  $\tilde S^\diamond_{j-m+1,j}$, $\tilde S^\diamond_{j-m+1}$, respectively.

Then by Lemma 4 of \cite{zhou2010simultaneous}, we have   uniformly with resepct to $t\in[0,1]$,
\begin{align}
|\E(\Lambda^\diamond_s(t))-\E(\Lambda_s(t))|=O(\sqrt{m/n}), ~~s=1,2,3.
\end{align}
By Lemma 5 of \cite{zhou2010simultaneous}, it follows for $s=1,2$,
\begin{align}
|\E(\Lambda^\diamond_s(t))-\sigma^2(t)/2|=O(m^{-1}+\tau_n^2), t\in[\gamma_n,1-\gamma_n],\\
|\E(\Lambda^\diamond_s(t))-\sigma^2(t)/2|=O(m^{-1}+\tau_n), t\in[0,\gamma_n)\cup (1-\gamma_n,1].
\end{align}
Define $\Gamma(k)=\E(G( {i/n},\FF_0)G( {i/n},\FF_k))$, then   similar arguments as given in the proof of   Lemma 5 of \cite{zhou2010simultaneous} yield $\Gamma(k)=O(\chi^{|k|})$. 
Elementary calculations show that for $1\leq j\leq n$
\begin{align}
\E(S_{j-m+1,j}^\diamond S_{j+1,j+m}^\diamond)=\sum_{k=1}^m\Gamma(k)=O(1),
\end{align}
which proves  
\begin{align}
\E(\Lambda^\diamond_3(t))=O(m^{-1})\label{2018-6.13}
\end{align}
uniformly with respect to $t\in[0,1]$.
From \eqref{2018-6.8}--\eqref{2018-6.13} it follows that 
\begin{align}
\sup_{t\in[\gamma_n,1-\gamma_n]}|\E\tilde \sigma^2(t)-\sigma^2(t)|=O(\sqrt{m/n}+m^{-1}+\tau_n^2),\\
\sup_{t\in[0,\gamma_n)\cup(1-\gamma_n,1]}|\E\tilde \sigma^2(t)-\sigma^2(t)|=O(\sqrt{m/n}+m^{-1}+ {\tau_n}) {.}
\end{align}
The theorem is now a consequence of these two equations and \eqref{2018-6.6}--\eqref{2018-6.8}.\hfill $\Box$

\section{Some technical results} \label{sec8}
\def\theequation{B.\arabic{equation}}
\setcounter{equation}{0}

\subsection{The size of mass excess}
\label{proofmainlemma}

\begin{lemma}\label{sizeset}
	Assume that the function $\mu(\cdot)-m$ has $k$ roots $0<t_1< \ldots <t_k<1$ of order $v_i$,
	$1\leq i\leq k$,
	and define $\gamma = \frac{1}{2}\min_{0\leq i\leq k}(t_{i+1}-t_i)$ (with convention that $t_0=0,t_{k+1}=1$), such that
	\begin{description}
		\item (i) 
		For $1\leq s\leq k$, the $(v_s+1)$nd derivative of $\mu(\cdot)$  is Lipschitz continuous on the interval $\mathcal I_s:=[t_s-\gamma,t_s+\gamma]$.
		\item (ii) $\mu(\cdot)$ is strictly monotone on the intervals  $ {\mathcal{I}}_s^-$ and ${ \mathcal{ I}}_s^+$   for $1\leq s\leq k$, where $\mathcal I_s^-:=[t_s-\gamma,t_s]$, $\mathcal I_s^+:=(t_s,t_s+\gamma]$,
		\item (iii) 
		there exists a positive number $\epsilon$, such that $\min_{t\in [0,1]\cap_{s=1}^k \bar{ \mathcal I}_s}|\mu(t)-m|\geq \epsilon$, where $\bar {\mathcal I}_s:=[0,t_s-\gamma)\cup(t_s+\gamma ,1]$ is
		complement of $\mathcal I_s$.
	\end{description}  
	If $A_n$ denotes the set
	\begin{align}\label{Nov-27-5.53}
	A_n:=\left\{s:|\mu(s)-m|\leq h_n\right\} {,}
	\end{align} 
	then there exists a sufficiently large constant $C$ such that for any sequence $h_n\rightarrow 0$, we have
	\begin{align}
	\lambda(A_n)\leq C h_n^{\frac{1}{v+1}},
	\end{align}
	where $v=\max_{1\leq l\leq k}v_l$.
	Furthermore, there exists a sufficiently large constant $M$, such that 
	\begin{equation} \label{part1}
	A_n=\cup_{l=1}^k B_{n,l,M}
	\end{equation} 
	when $n$ is sufficiently large, where the sets $B_{n,l,M}$ are defined by
	\begin{align}
	B_{n,l,M}=\{s: |s-t_l|^{v_l+1}\leq M h_n, |\mu(s)-m|\leq h_n\}.
	\end{align}
	
\end{lemma}

\noindent
{\it Proof.} 
Define for $1\leq l\leq k$, \begin{align}\label{important}
A_{n,l}=\left\{s:|\mu(s)-m|\leq h_n, |s-t_l|< \min\{\gamma,\zeta_n\}\right\},
\end{align}
where $\zeta_n$ is a sequence of real numbers which converges to zero arbitrarily slowly.
We shall show that there exists a constant $n_0 \in \mathbb{N}$, such that for $n \geq n_0$
\begin{align}
A_n &=\cup_{l=1}^k A_{n,l},
\label{subi} \\
A_{n,l} &\subseteq B_{n,l,M}, \quad 1\leq l\leq k, \label{auxillary}
\end{align}
where $M$ is a sufficiently large constant.
Note that \eqref{subi} and \eqref{auxillary} yield $A_n\subseteq \cup_{l=1}^k B_{n,l,M}$. By definition  of $B_{n,l,M}$ and $A_n$, we have that
$\cup_{l=1}^k B_{n,l,M}\subseteq A_n$, which  proves \eqref{part1}.
Then a straightforward calculation shows that $$\lambda(B_{n,l,M})\leq C h_n^{\frac{1}{v_l+1}}\leq C h_n^{\frac{1}{v+1}},$$ and the lemma follows.

\noindent We first prove the assertion \eqref{subi}.
By definition, $A_n\supseteq \cup_{l=1}^k A_{n,l}$. We now argue that there exists a sufficiently large constant $n_0$, such that for $n\geq n_0$, $\cup_{l=1}^k A_{n,l}\supseteq A_n$. 

Suppose this statement is not true, then  there exists a sequence of points $(s_n)_{n \in \mathbb{N}}$,  such that $s_n\in A_n$ and $s_n\in \cap_{l=1}^k \bar A_{n,l}$, where $\bar A_{n,l}$ is the complement set of $A_{n,l}$. Since $h_n=o(1)$ we have  $h_n<\epsilon$ for sufficiently large $n$ and by assumption (iii),  there exists an $l\in \{1, \dots ,k\}$ such that
$$s_n \in {\mathcal I}_l\cap A_n\cap \bar A_{n,l}.$$
Without loss of generality  we assume   that $s_n\in {\mathcal I}^+_l\cap A_n\cap\bar A_{n,l}.$ The case that $s_n\in {\mathcal I}^-_l\cap A_n\cap\bar A_{n,l}$ can be treated similarly.

A Taylor expansion and assumption (i)  yield for   sufficiently large $n \in \mathbb{N}$ 
\begin{align}\mu(s)-\mu(t_l)=\frac{\mu^{(v_l+1)}(t_l)}{(v_l+1)!}(s-t_l)^{v_l+1}+\frac{\mu^{(v_l+1)}(t_l^*)-\mu^{(v_l+1)}(t_l)}{(v_l+1)!}(s-t_l)^{v_l+1}\label{localtylor}
\end{align}
for $s\in A_{n,l}$,
where $t^*_l\in [t_l\wedge s,t_l\vee s]$. By the definition of $A_{n,l}$ in \eqref{important} and the fact that $\zeta_n=o(1)$,  we have that $A_{n,l}\subset \mathcal I_l$ for sufficiently large $n \in \mathbb{N}$. This result together with $s_n\in {\mathcal I}^+_l\cap A_n\cap\bar A_{n,l}$ implies that $t_l+\zeta_n<s_n\leq t_l+\gamma$. However, by assumption (ii), $\mu(\cdot)$ is strictly monotone in $\mathcal I^+_l$, which yields that for sufficiently large $n$, 
\begin{align}
|\mu(s_n)-\mu(t_l)|\geq |\mu(t_l+\zeta_n)-\mu(t_l)|>2h_n,\label{contradition1}
\end{align}
where the last $>$ is due to \eqref{localtylor}, the Lipschitz continuity of $\mu^{(v_l+1)}(\cdot)$ in the neighbourhood of $t_l$ and the fact that $\zeta_n\rightarrow 0$ arbitrarily slowly. By the definition of $A_n$ in \eqref{Nov-27-5.53}, equation \eqref{contradition1} implies that $s_n\not \in A_n$. This contradicts to the assumption that
$s_n\in A_n$, from which \eqref{subi} follows.

Now we show the conclusion \eqref{auxillary}.  Since $\mu(t_l)=m$ and the leading term in \eqref{localtylor} is of order $|(s-t_l)^{v_l+1}|$, the set $A_{n,l}$ can be represented as
\begin{align}
\Big\{s:|s-t_l|\leq \Big(\frac{h_n}{|M_{1,l}+M_{2,l}(s)|}\Big)^{\frac{1}{v_l+1}}, |s-t_l|\leq \zeta_n,|\mu(s)-c|\leq h_n\Big\},
\end{align}
where $M_{1,l}=\frac{\mu^{(v_l+1)}(t_l)}{(v_l+1)!}$, and $M_{2,l}(s)=\frac{\mu^{(v_l+1)}(t_l^*)-\mu^{(v_l+1)}(t_l)}{(v_l+1)!}(s-t_l)^{v_l+1}$ for some  $t_l^*\in[t_l\wedge s,t_l\vee s]$.  By the Lipschitz continuity of $\mu^{(v_l+1)}(\cdot)$ on the interval $[t_l-\gamma, t_l+\gamma]$, there exists a constant $M_l'$ such that $|M_{2,l}(s)|\leq M_l'|t_l-s|$. 
As $\zeta_n=o(1)$ there exists  an $n_l\in \mathbb N$ such that  $|s-t_l|\leq \frac{|M_{1,l}|}{2M_l'} $ for all $s\in A_{n,l}$ whenever $n \geq n_l$. This yields  $$\left|M_{1,l}+M_{2,l}(s-t_l)\right|\geq \frac{|M_{1,l}|}{2}$$
for all $n\geq n_l$, $s\in A_{n,l}$.  By choosing $n_0=\max_{1\leq l\leq k}n_l$ and $M=\max_{1\leq l\leq k}\left(\frac{2}{|M_{1,l}|}\right)^{\frac{1}{v_l+1}}$, and noticing the fact that $\zeta_n\rightarrow 0$ arbitrarily slow, it follows that $$A_{n,l}\subseteq B_{n,l,M}$$ for $n\geq n_0.$
Thus (\ref{auxillary}) follows, which completes the proof of Lemma \ref{sizeset}. \hfill $\Box$

\begin{remark}\label{disjoint}  {
		{\rm	Observe  that $B_{n,i,M}\cap B_{n,j,M}=\emptyset$ for $i\neq j$ if $n$ is sufficiently large. Moreover, $B_{n,i,M}$ can
			be covered by closed intervals. The Lemma shows that the set $\{t:|\mu(t)-m|\leq h_n,t\in [0,1]\}$ can be
			decomposed in disjoint  intervals containing the root of the equation $\mu(t)=m$, with   Lebesgue measure   determined by the maximal critical order of the  roots.}
	}
\end{remark}

\subsection{Uniform bounds for nonparametric estimates} \label{proofstechnical1}

In this section we present some results about the rate of uniform convergence of  the Jackknife estimator
$\tilde \mu_{b_n}(t)$   defined in \eqref{Jack}. 

\begin{lemma}\label{Jackknife Approximation}
	Recall the definition of   $\tilde \mu_{b_n}$ in \eqref{Jack} and
	suppose that  Assumption \ref{A1}(a) holds. If  $b_n\rightarrow 0$, $nb_n\rightarrow \infty$, then \begin{align}
	\sup_{t\in [b_n,1-b_n]}\Big|\tilde \mu_{b_n}(t)-\mu(t)-\frac{1}{nb_n}\sum_{i=1}^nK^*\Big(\frac{i/n-t}{b_n}\Big)e_i\Big|=O(b_n^{3}+\frac{1}{nb_n}), \label{Appendix.5.63}\\
	\Big|\tilde \mu_{b_n}(0)-\mu(0)-\frac{1}{nb_n}\sum_{i=1}^n\bar K^*\Big(\frac{i/n}{b_n}\Big)e_i\Big|=O(b_n^{3}+\frac{1}{nb_n}),
	\end{align}
	where  $K^*(\cdot)$ and $\bar K^*(\cdot)$ are defined  in \eqref{k1} and \eqref{k2}, respectively 
\end{lemma}

\noindent
{\it Proof.} We only show the estimate \eqref{Appendix.5.63}. The other result follows similarly   using Lemma B.2 of \cite{dette2015change}. By Lemma B.1 of \cite{dette2015change}
we obtain a uniform bound for the (uncorrected) local linear
estimate $\hat \mu_{b_n}$ in \eqref{Locallinear}, that is
\begin{align}\label{Appendix_mu}
\sup_{t\in [b_n, 1-b_n]} \Bigl |\hat{\mu}_{b_n}(t)-\mu(t)-\frac{\mu_2\ddot{\mu}(t)}{2}b_n^2-\frac{1}{nb_n}\sum_{i=1}^ne_iK_{b_n}(i/n-t)\Bigr |=O(b_n^3+\frac{1}{nb_n}).
\end{align}
Then the lemma follows from the definition of $\tilde \mu_{b_n}(\cdot)$. \hfill $\Box$

\bigskip

\begin{lemma}\label{maxdevvar1}
	If   Assumption \ref{A1}(a), Assumption \ref{A2}
	are satisfied and  $\frac{nb^2_n}{\log ^4 n}\rightarrow \infty$, $b_n\rightarrow 0$, then
	\begin{align}
	\sup_{t\in\{0\} \cup [b_n,1-b_n]}|\tilde \mu_{b_n}(t)-\mu(t)|
	=O_p\Big(b_n^3+\frac{\log n}{\sqrt{nb_n}}\Big).\label{New.16}\\
	\sup_{t\in [0,b_n)\cup(1-b_n,1 {]}}|\tilde \mu_{b_n}(t)-\mu(t)|
	=O_p\Big(b_n^2+\frac{\log n}{\sqrt{nb_n}}\Big).\label{New.17}
	\end{align}
\end{lemma}
\noindent
{\it Proof.} We only  prove the estimate
$$\sup_{ [b_n,1-b_n]}|\tilde \mu_{b_n}(t)-\mu(t)|
=O_p\Big(b_n^3+\frac{\log n}{\sqrt{nb_n}}\Big).$$ The case that $t=0$
in \eqref{New.16} and the estimate \eqref{New.17} follow by  similar arguments, which are omitted for the sake of brevity. By
the stochastic expansion \eqref{Appendix.5.63}, it suffices to show that
\begin{align} \sup_{t\in [b_n,1-b_n]}\Big|\frac{1}{nb_n}\sum_{i=1}^nK^*\Big(\frac{i/n-t}{b_n}\Big)e_i\Big|=O_p\Big(\frac{\log n}{\sqrt{nb_n}}\Big).
\end{align}
Then Assumption \ref{A2}, Proposition 5 of \cite{zhou2013} and the summation by parts formula  (44) in \cite{zhou2010nonparametric} yield the existence (on a possibly richer probability space) of a sequence  $(V_i)_{i\in \mathbb Z}$ of independently standard normal distributed random variables such that
\begin{align} \sup_{t\in [b_n,1-b_n]}\Big|\frac{1}{nb_n}\sum_{i=1}^nK^*\Big(\frac{i/n-t}{b_n}\Big)(e_i-V_i)\Big|=O_p\Big(\frac{n^{1/4}\log^2 n}{nb_n}\Big).
\end{align}
Note that $(V_i)_{i\in \mathbb Z}$  is a martingale difference sequence with respect to the filtration generated by $(V_{-\infty},...,V_i)$.
By Burkholder's inequality it follows that for any positive $\kappa$ and a sufficiently large universal constant $C$ the inequality
\begin{align}
& \Big\|\sum_{i=1}^nV_iK^*_{b_n}(i/n-t)\Big \|^2_\kappa\leq C \kappa \Big\|
\big( {\sum_{i=1}^n\{V_iK^*_{b_n}(i/n-t)\}^2} \big)^{1/2} \Big \|_\kappa^2\notag
\\&\leq C\kappa\sum_{i=1}^n \big \|(V_iK^*_{b_n}(i/n-t))^2 \big \|_\frac{\kappa}{2}\notag =C\kappa \sum_{i=1}^n \big \|(V_iK^*_{b_n}(i/n-t)) \big \|_\kappa^2\leq C\kappa^2(nb_n)
\end{align}
holds uniformly with respect to $t\in[b_n,1-b_n]$,
where  we have used that $\E|V_0|^\kappa\leq (\kappa-1)!!\leq \kappa^{\frac{\kappa}{2}} $
in the last inequality. This leads to
\begin{align*}
\sup_{t\in [b_n,1-b_n]}\Big\|\frac{1}{nb_n}\sum_{i=1}^nK^*_{b_n}(i/n-t)V_i\Big\|_\kappa=O\Big(\frac{\kappa}{\sqrt{nb_n}}\Big).
\end{align*}
Similarly, we obtain
\begin{align*}
\sup_{t\in [b_n,1-b_n]}\Big\|\frac{1}{nb_n}\sum_{i=1}^n\frac{\partial}{\partial t}K^*_{b_n}(i/n-t)V_i\Big\|_\kappa=O\Big(\frac{\kappa b_n^{-1}}{\sqrt{nb_n}}\Big).
\end{align*}
Consequently, Proposition B.1. of \cite{dette2015change}  shows that
\begin{align*}\Big\| \sup_{t\in [b_n,1-b_n]}\frac{1}{nb_n}\sum_{i=1}^nK^*_{b_n}(i/n-t)V_i\Big\|_\kappa=O\Big(\frac{\kappa b_n^{-\frac{1}{\kappa}}}{\sqrt{nb_n}}\Big)
\end{align*}
The result now follows using  $\kappa=\log (b_n^{-1})$ observing the conditions on the  bandwidths.

 \end{document}